\documentclass[aps,pra,reprint,groupedaddress,showpacs,showkeys]{revtex4-1}

\usepackage{watermark}
 \usepackage{amsmath}
 \usepackage{amsfonts}
 \usepackage{graphicx}
 \usepackage{epsfig}
 \usepackage{epstopdf}

 \usepackage{hyperref}
 \hypersetup{
    bookmarks=true,         
    unicode=false,          
    pdftoolbar=true,        
    pdfmenubar=true,        
    pdffitwindow=false,     
    pdfstartview={FitH},    
    pdftitle={Vertex enhancement for positron annihilation with core electrons},    
    pdfauthor={D. G. Green},     
    pdfsubject={Positron_hlikeions},   
    pdfcreator={D. G. Green},   
    pdfproducer={Producer}, 
    pdfkeywords={keywords}, 
    pdfnewwindow=true,      
    colorlinks=true,       
    linkcolor=blue,          
    citecolor=blue,        
    filecolor=magenta,      
    urlcolor=blue           
}

\newcommand{\au}{\,a.u.}

\newcommand{\etal}{\emph{et al.}}
\newcommand{\abin}{\emph{ab initio}}
\newcommand{\eqn}[1]{Eq.\,\ref{#1}}
\newcommand{\fig}[1]{Fig.\,\ref{#1}}
\newcommand{\tab}[1]{Table~\ref{#1}}
\newcommand{\secref}[1]{Sec.\,\ref{#1}}

\begin{document}
\title{Positron annihilation with core and valence electrons}
\author{D.~G. Green}
\email[Correspondences to:~]{dermot.green@balliol.oxon.org}
\altaffiliation{\newline Present address: {Joint Quantum Centre (JQC) Durham/ \hspace{-0.7ex}Newcastle}, Departments of Chemistry \& Physics, Durham University, South Road, Durham, DH1 3LE, UK.}
\author{G.~F. Gribakin}
\email{g.gribakin@qub.ac.uk}
\affiliation{Department of Applied Mathematics and Theoretical Physics, 
Queen's University Belfast, Belfast, BT7\,1NN, Northern Ireland, United Kingdom}
\date{\today}

\begin{abstract}
$\gamma$-ray spectra for positron annihilation with the core and valence electrons of the noble gas atoms Ar, Kr and Xe is calculated within the framework of diagrammatic many-body theory.
The effect of positron-atom and short-range positron-electron correlations on the annihilation process is examined in detail.
Short-range correlations, which are described through non-local corrections to the vertex of the annihilation amplitude, are found to significantly enhance the spectra for annihilation on the core orbitals. 
For Ar, Kr and Xe, the core contributions to the annihilation rate are found to be 0.55\%, 1.5\% and 2.2\% respectively, their small values reflecting the difficulty for the positron to probe distances close to the nucleus. Importantly however, the core subshells have a broad momentum distribution and markedly contribute to the annihilation spectra at Doppler energy shifts $\gtrsim3$\,keV, and even dominate the spectra of Kr and Xe at shifts $\gtrsim5$\,keV.
Their inclusion brings the theoretical spectra into excellent agreement with the experimental $\gamma$-spectra across the full range of Doppler energy shifts.
Additionally, the theory enables the calculation of the `exact' vertex enhancement factors $\bar{\gamma}_{n\ell}$ for individual core and valence subshells $n\ell$.
They are found to follow a simple and physically motivated scaling with the subshell ionization energy $I_{n\ell}$: $\bar{\gamma}_{n\ell}=1+\sqrt{A/I_{n\ell}}+(B/I_{n\ell})^{\beta}$, where $A$, $B$ and $\beta$ are positive constants. 
It is demonstrated that these factors can be incorporated in simple independent-particle-model calculations to successfully reconstruct the true many-body annihilation $\gamma$-spectra. 
This formula can be used to determine the enhancement factors for positron annihilation with core electrons of atoms across the periodic table, and with the localized atomic-like core electrons of condensed matter systems.
\end{abstract}
\maketitle

\section{Introduction}

Low-energy positrons annihilate predominantly on the outermost valence electrons in atoms, molecules and in condensed matter systems \cite{PhysRevA.55.3586,RevModPhys.66.841}. 
Small fractions of positrons can, however, tunnel through the repulsive nuclear potential and annihilate on the core electrons \footnote{Core electrons, likewise, tunnel through the atomic binding potential towards the positron, and their wave functions are affected by ``exchange-assisted tunneling'' \cite{PhysRevA.87.042511}.}.
The $\gamma$-rays produced in the annihilation events give rise to a Doppler-broadened energy spectrum that is characteristic of the electronic states involved, e.g., annihilation on the tightly-bound core electrons results in a distinct signature in the high-energy region of the spectrum.
A dramatic r\^ole is played in the annihilation process by short-range electron-positron and positron-atom correlations: they significantly enhance annihilation rates \cite{PhysRevA.51.473, DGG_posnobles} and alter the shape and magnitude of the annihilation $\gamma$-spectra \cite{0953-4075-39-7-008, DGG_hlike, DGG_molgammashort, DGG_molgamma, DGG_coreprl}. 

In this work we examine, systematically and in detail, the effect of these correlations on the $\gamma$-spectra for positrons annihilating on individual core and valence subshells in many-electron atoms. 
Specifically, we use many-body theory \cite{PhysRevA.70.032720,0953-4075-39-7-008,DGG_hlike,DGG_posnobles} (MBT) to calculate the annihilation $\gamma$-ray spectra and annihilation rates for positron annihilation on individual subshells of the noble gases Ar, Kr and Xe. 
We show that the theory is the first to accurately reproduce the the measured spectra \cite{PhysRevA.55.3586} (see Ref. \cite{DGG_coreprl} for a brief account of this part of the work). Furthermore, we establish firmly the fraction annihilation on individual subshells, and uncover a simple scaling of the correlation enhancement factors, which describe the increase in the annihilation probability beyond the independent-particle approximation (IPA), with the electron ionization energy \cite{DGG_coreprl}.
This work, taken together with our work in Refs. \cite{DGG_posnobles} and \cite{DGG_coreprl}, gives a (near) complete description of the positron noble-gas system.  

The ability to discern the core signal in the $\gamma$-spectrum was first realised 
experimentally in \cite{PhysRevLett.38.241}. 
The elemental specificity arising from the chemically distinct signature in the large Doppler-shift regions of the spectra corresponding to core annihilation has since been exploited \cite{PhysRevLett.77.2097,PhysRevLett.82.3819,PhysRevB.51.4176}, and measurement of the signal of individual selected core levels has been demonstrated \cite{PhysRevLett.89.075503}.
Positron annihilation on core electrons is now a key process in several established experimental techniques, e.g., positron-induced Auger-electron spectroscopy (PAES) \cite{PhysRevLett.61.2245,Ohdaira1997177,nepomucpaes,Weiss2007285},  used for diagnostics of industrially important materials, and the recently emerged time-resolved PAES \cite{nepomucref}, to study the dynamics of catalysis, corrosion, and surface alloying \cite{PhysRevLett.105.207401}. Coincident measurements of the $\gamma $-rays and Auger electrons yields $\gamma $-ray spectra for individual core orbitals \cite{PhysRevLett.89.075503,PhysRevB.73.014114}.

In all cases, interpretation of the measured annihilation $\gamma$-spectra relies heavily on theoretical input.
For example, to interpret the results of PAES one needs to know the relative probabilities of 
positron annihilation with inner electrons of various atoms \cite{Weiss2007285}. 
These are often computed using the independent-particle approximation (IPA) \cite{PhysRevLett.38.241,PhysRevB.41.3928}, 
or in a modified IPA framework that includes phenomenological enhancement factors that attempt to account for the important effects of electron-positron correlations \cite{JPhysCM.1.10595,PhysRevB.51.4176,PhysRevB.54.2397, PhysRevB.43.2580, JPhysCM.3.7631, PhysRevLett.89.075503, PhysRevB.73.014114, Weiss2007285,PhysRevB.57.12219}. 
In most cases, the materials of interest are condensed matter systems, and the enhancement factors are usually calculated using a variety of density functional theory methods, e.g., the local density approximation (LDA) \cite{RevModPhys.66.841,PhysRevB.34.3820}, the generalized gradient approximation (GGA) \cite{PhysRevB.51.7341,PhysRevB.53.16201,Barbiellini1997283,PhysRevB.56.7136}, or the weighted density approximation (WDA) \cite{PhysRevB.44.10857,JPhysCM.5.8195,PhysRevB.58.11285}, all of which rely heavily on theoretical considerations of the electron gas. 
However, due to the strong variations in the density, the LDA is not expected to work well for the core electrons, and has been found to overestimate the annihilation rates \cite{PhysRevB.58.11285}
The enhancement factor approach within the one-component and Boro\ifmmode \acute{n}\else \'{n}\fi{}ski-Nieminen two-component LDA \cite{PhysRevB.34.3820}, as well as within the GGA, have been tested and compared with bound positron-atom stochastic variational calculations in Ref. \cite{PhysRevB.65.235103}, although no conclusive statements were made regarding the core enhancements.

One particular weakness of the DFT enhancement factor approach is that it obscures the true non-local many-body nature of the problem.
The early theoretical works of Kahana, Carbotte and others (see, e.g., Refs. \cite{PhysRev.117.123,PhysRev.129.1622,PhysRev.144.309,PhysRev.155.197,PhysRev.188.550,PhysRev.167.239,PhysRevB.20.883,PhysRevB.12.1689,PhysRev.136.1728,PhysRev.150.243,chiba1977}) on positron annihilation were based on many-body theory, which inherently incorporates the non-local effects.
Naturally however, these methods contained numerous approximations, and again were mainly annihilation rate calculations for the electron gas.
Some effects of the non-local components of the amplitude have been included in phenomenological enhancement factor calculations using the WDA \cite{PhysRevB.58.11285}, and to a lesser extent in the GGA \cite{PhysRevB.51.4176,PhysRevB.54.2397}. However, these methods again rely on considerations of the electron gas.

Regarding the core enhancement factors in particular, there are a number of calculations that differ in their treatment and overall, there is no definite agreement in their values.
At one extreme it is assumed that the core enhancement is negligible, while at the other it is assumed to be equivalent to the valence enhancement (see, e.g., \cite{PhysRevB.58.11285} and references therein). 
In reality, the core electrons are less susceptible to a perturbation from the interaction with the incident positron than the valence ones. 
One should therefore expect the core spectra to be enhanced by a factor 1$<\bar\gamma_{\rm core}<$$\bar\gamma_{\rm valence}$, i.e., one should expect the correlational enhancement to be weaker for the tightly-bound core than for the valence electrons.
Indeed, a recent study by the authors that investigated the correlational enhancement for annihilation of positrons on H-like ions, for which the electron binding energies span a range typical of those of core electrons in many-electron atoms, suggested that the core enhancement could be significantly greater than unity \cite{DGG_hlike}.

For positron annihilation in condensed matter systems, an alternative approach to the calculation of core electron enhancement factors is suggested by the fact that the core electrons of these systems are localized and atomic in nature. 
Their enhancement factors are therefore open to \abin\ atomic-based calculations. 
This was realised by Bonderup \emph{et al.} \cite{PhysRevB.20.883}, who, in order to elucidate the r\^ole of electron-positron correlations in condensed matter systems,  made a first-order perturbation theory estimate of the enhancement for positron annihilation on hydrogen-like ions.
Compared to condensed matter systems, the problem of positrons annihilating in atomic systems is in principle, much simpler.
For one, simplifications result from the spherical symmetry of the atom that is not present in the condensed matter systems.
For the latter, increased complexity comes from the periodic crystal potential: calculating the electronic and band structure is much more involved, and from the Bloch-wave nature of the incident positron wave function, which must be properly accounted for.
Despite this, for annihilation on core electrons, positron-condensed-matter studies are much more numerous than corresponding positron-atom studies. 

Evidence of the core contribution to the annihilation $\gamma$-spectra of the noble gases was first reported by Iwata \etal \cite{PhysRevA.55.3586}, 
in which upper bounds (2\% and 3\% for Ar and Kr respectively) were set on the core contribution to the total spectra. 
This was followed by the first experimental study of 
positrons annihilating with inner-shell electrons in isolated two-body positron-atom interactions \cite{PhysRevLett.79.39}, in which more precise 
measurements of the $\gamma$-spectra of the noble gases Ar, Kr and Xe were reported.
In that work, a static Hartree-Fock calculation was successfully used to identify the inner-shell electron contribution to the measured spectra. 
However, owing to the fact that it ignored the state dependent enhancement, the relative fraction of annihilations with inner-shell electrons was not predicted accurately by this theory.
Almost a decade later, Dunlop and Gribakin \cite{0953-4075-39-7-008} used many-body theory to explain the main features of the $\gamma$-spectra of the noble gases. 
In that work, however, only annihilation on the valence shells was considered, the short-range electron-positron interaction was treated only in first-order, and positron-atom correlations were ignored; significant discrepancies between the theoretical results and experiment were evident.
Dunlop \cite{Dunlop_thesis} later attempted to calculate the core contribution to the spectra via a semi-empirical fitting procedure.
For this, the annihilation on the valence shells was calculated with the full incorporation of short-range electron-positron correlations, whereas the core was treated in the simple independent-particle-model (IPA) approximation.
It will be shown however, that neglecting the core enhancement in this way leads to an overestimate of the core contribution to the $\gamma$-spectrum. 

The purpose of this work is three-fold:  (1) to examine in detail the effect of electron-positron correlations on the process of annihilation on tightly-bound core electrons of many-electron atoms (using the noble gases as test cases) 
and in doing so; (2) fully explain the measured spectra of the noble gases and establish the fraction of annihilation on individual core and valence subshells,
and finally;
(3) to calculate the `\emph{true}' core-electron enhancement factors (and demonstrate that they can be used to correct relatively simple independent-particle-approximation calculations to produce accurate spectra).

The structure of the paper is as follows. 
In Sec.~\ref{sec:theory} we outline the many-body
theory of annihilation $\gamma$-spectra, detailing the form of the fully 
correlated incident positron wave function and annihilation vertex.  In Sec.~\ref{sec:numerics} 
we briefly discuss the numerical implementation of the theory. 
In Sec.~\ref{sec:results} we present our calculated annihilation spectra for Ar, Kr and Xe, comparing them with the experimental results of Iwata \emph{et al.} \cite{PhysRevLett.79.39}.
In addition, the partial $Z_{\rm eff}$ are calculated for each subshell.
In the final part, we calculate vertex enhancement factors and demonstrate how they can be used in independent-particle-model calculations to obtain accurate annihilation spectra. 
A summary is given in Sec.~\ref{sec:conclusions}. 

\section{Many-body theory of annihilation $\gamma$-spectra}\label{sec:theory}

\subsection{Basics}
The process of electron-positron annihilation is described fundamentally by 
quantum electrodynamics \cite{Akhiezer,QED}.
It dictates that low-energy positron annihilation proceeds predominately via two photon production, 
a process in which the total spin of the electron-positron system must be zero\;\cite{Akhiezer,QED}.
Consider then a positron with momentum $\mathbf k$ annihilating in a many-electron system with an electron in state $n$\,\footnote{Here $n$ refers to a general set of quantum numbers. In this work it will usually refer to the principal quantum number $n$ as well as the angular momentum quantum number $l$ of the state}.
In the annihilation event, two gamma-photons with total momentum 
${\bf p}_{\gamma_1}+{\bf p}_{\gamma_2}=\mathbf P$ are produced. 
Let us denote the amplitude for this process by
\begin{eqnarray}
A_{n{\bf k}}({\bf P})= \langle \Psi_{n}^{N-1};2\gamma{\bf P}|\Psi_{\bf k}^{N+1}\rangle\label{eqn:annamp},
\end{eqnarray}
where $\Psi_{\bf k}^{N+1}({\bf r}_1,\dots{\bf r}_N; {\bf r})$ is the true, fully-correlated many-particle wave function of the initial state of $N$ 
electrons (coordinates ${\bf r}_i$) and the positron (coordinate ${\bf r}$), and $\Psi_{n}^{N-1}({\bf r}_1,\dots{\bf r}_{N-1})$ is that of the final state of system with $N-1$ electrons and a hole in state $n$. 
At large positron-atom separations the initial $N+1$ particle wave function has the form
\begin{eqnarray}
\Psi_{\bf k}^{N+1}({\bf r}_1,\dots{\bf r}_N; {\bf r})\simeq \Psi^N_0({\bf r}_1,\dots{\bf r}_N) e^{i{\bf k}\cdot{\bf r}}
\end{eqnarray}
which is normalized to the positron plane wave, where $\Psi^N_0$ is the atomic ground state wave function.
In the centre-of-mass frame, where the total momentum ${\bf P}$ is zero, the two photons 
propagate in opposite directions and have equal energies 
$E_{\gamma}=p_{\gamma}c=mc^2+\frac{1}{2}(E_{\rm i}-E_{\rm f})\simeq mc^2\simeq 511\,{\rm keV}$,
where $E_{\rm i}$ and $E_{\rm f}$ denote the energy of the initial and final states (excluding rest mass). 
When ${\bf P}$ is non-zero however, the two photons no longer propagate in exactly opposite directions
and their energy is Doppler shifted. For example, for the first photon
$E_{\gamma_1}=E_{\gamma}+mcV\cos{\theta}$, where ${\bf V}={\bf P}/2m$ is the centre-of-mass
velocity of the electron-positron pair and $\theta$ is the angle between the direction of the photon velocity ${\bf c}$
and $\bf{V}$.  
Assuming that $V\ll c$, and $p_{\gamma_1}=E_{\gamma_1}/c\approx mc$, the shift 
of the photon energy from the centre of the line, $\epsilon=E_{\gamma_1}-E_{\gamma}$, is
\begin{eqnarray}
\epsilon=mc\;V\cos{\theta}=\frac{Pc}{2}\cos{\theta}=\frac{1}{2}{\bf P}\cdot{\bf c}.
\end{eqnarray}
The typical momenta of electrons bound with energy $\varepsilon_n$ determine the Doppler width
 of the annihilation spectrum, $\epsilon \sim Pc \sim\sqrt{|\varepsilon_n|mc^2}\gg|\varepsilon_n|$. 
Hence the shift of the line centre $\varepsilon_n/2$ from $E_\gamma=mc^2=511$\,keV
can usually be neglected, even for the core electrons.
The annihilation $\gamma$-ray (Doppler) spectrum can then be written in a form similar to a Compton profile, 
as the projection of the annihilation probability:
\begin{eqnarray}
w_{n{\bf k}}(\epsilon,{\bf c})=\int |A_{n{\bf k}}({\bf P})|^2 \delta\left(\epsilon-\frac{1}{2}{\bf P}\cdot{\bf c}\right) \frac{d^3P}{(2\pi)^3}.
\end{eqnarray}
For gaseous systems, however, the experimentally accessible quantity is the average of this over 
the direction of emission of the annihilation photons ${\bf c}=c|\hat{\bf c}|$. 
Also, to compare with experiments that use positrons confined in a trap, one must also average over the direction of the incident positron momentum ${\bf k}$ giving 
\begin{eqnarray}\label{eqn:gammaspectra}
\bar{w}_{n\varepsilon}(\epsilon)&\equiv&\int w_{n{\bf k}}(\epsilon,{\bf c}) \frac{d\Omega_{{\bf c}}}{4\pi}\frac{d\Omega_{{\bf k}}}{4\pi},\\
&=&\frac{1}{c} \int _{2|\epsilon|/c}^{\infty} \int_{\Omega_{\bf P}} |A_{n\varepsilon}({\bf P})|^2 \frac{ d\Omega_{\bf P}}{(2\pi)^3} PdP
\end{eqnarray}
where 
$|A_{n\varepsilon}({\bf P})|^2\equiv\int |A_{n{\bf k}}({\bf P})|^2{d\Omega_{{\bf k}}}/{4\pi}$.
The annihilation amplitude $A({\bf P})$ is therefore central to the evaluation of the annihilation $\gamma$-spectra. Its calculation from many-body theory is outlined below. 
In most experiments, including those to which we compare our results, the annihilation photons are not detected in coincidence with the final state of the atom.
The observed spectrum is then given by a sum over the electronic states of the system: $\bar{w}_{\varepsilon}(\epsilon)=\sum_n\bar{w}_{n\varepsilon}(\epsilon)$. 
As we shall see, the total spectrum still retains distinct characteristics of the individual final states.

A useful quantity that is related to the $\gamma$-spectra is the normalized annihilation rate parameter $Z_{\rm eff}$. 
It is defined as the ratio of the true annihilation rate $\lambda$ to the spin-averaged Dirac annihilation rate on a free electron gas of number density $n_e$, $\bar\lambda_D=\pi r_0^2cn_e$ \cite{Fraser,pomeranchuk}
\begin{eqnarray}
\lambda=\bar\lambda_D Z_{\rm eff},
\end{eqnarray}
where $r_0=e^2/mc^2$ is the classical radius of the electron (in CGS units).
If the electrons belong to the atoms in a gas of number density $n$, then $n_e=Zn$, where $Z$ is the number of electrons per ion.
The corresponding spin-averaged annihilation cross section is
\begin{eqnarray}
\bar\sigma_{2\gamma}=\frac{\lambda}{n_ev}=\pi r_0^2\frac{c}{v} Z_{\rm eff}.
\end{eqnarray}
where $v$ is the positron velocity. 
By definition, $Z_{\rm eff}$ gives the \emph{effective number of electrons} that contribute to the annihilation. 
In this way it provides a measure of the strength of the correlations.
It can be determined from the annihilation $\gamma$-ray Doppler spectrum through the relation
\begin{eqnarray}\label{eqn:zeffspec}
Z_{{\rm eff}}=\int_{-\infty}^{\infty}  \bar{w}(\epsilon)\,d\epsilon.
\end{eqnarray} 
\begin{figure*}[!htb]
\includegraphics*[width=0.7\textwidth]{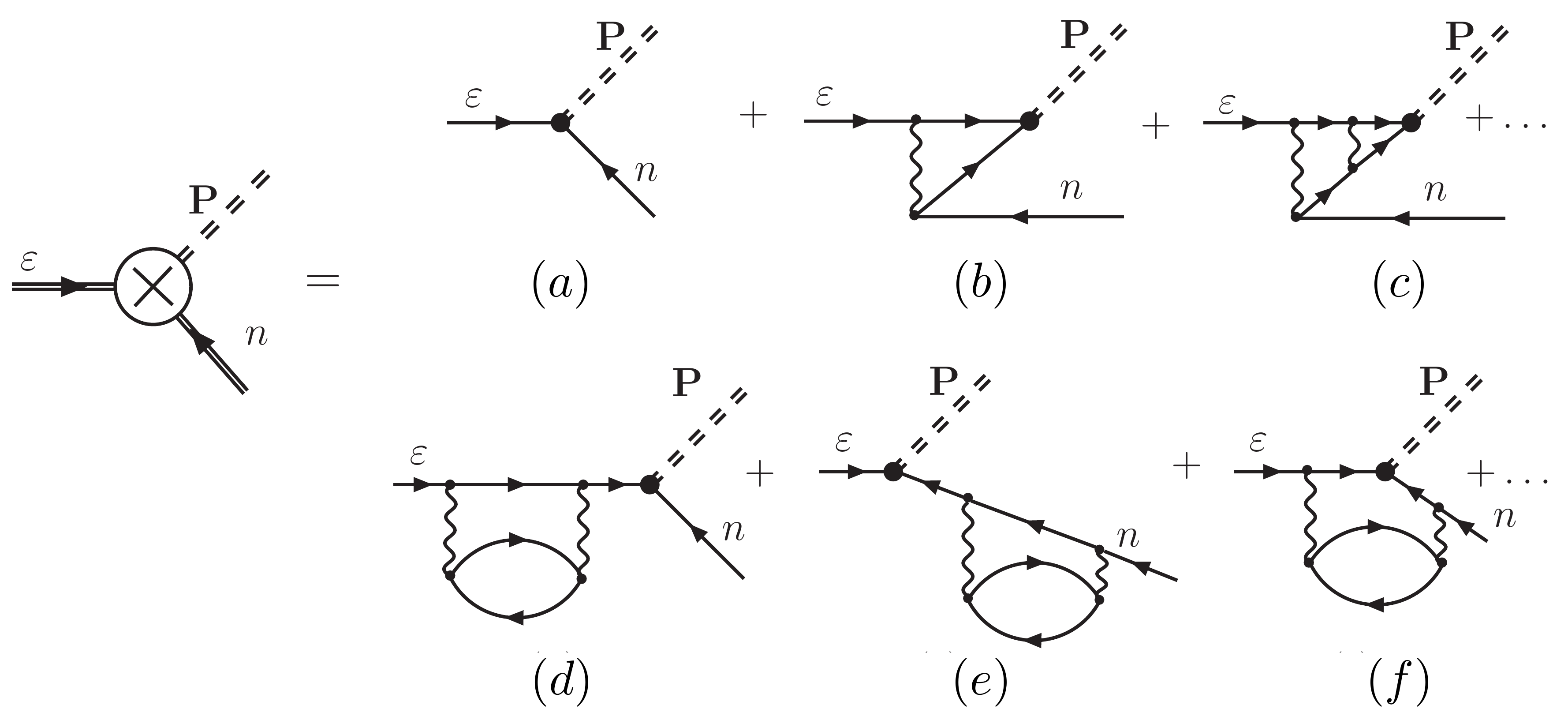}
\caption{Diagrammatic expansion of the annihilation amplitude $A_{n\varepsilon}({\bf P})$.
Diagram (a) represents the zeroth-order (independent-particle-model) approximation. 
All other diagrams are corrections to this that can be grouped into two independent types.
The first type, which includes diagrams (b), (c) and (f) are corrections to the annihilation vertex.
The second type, which includes diagrams (d) and (e) are corrections to the zeroth-order Hartree-Fock wave functions.
The general form of the amplitude, shown on the LHS of the equality, consists of four elements: (1) the incident positron wave function labelled by the quantum numbers $\varepsilon$ (the double line indicates the `dressed', i.e., fully correlated wave function); (2) the general annihilation vertex (crossed circle); (3) the wave function of the annihilated electron (hole) of state $n$, and finally; (4) the two $\gamma$-rays of total momentum ${\bf P}$.
\label{fig:anndiagsgen}}
\end{figure*}
The measurements to which we will compare our calculations are those of Surko and co-workers of the San Diego group \cite{PhysRevLett.79.39}. 
These experiments used a positron trap and low gas densities. 
In this regime, the parameter $Z_{\rm eff}$ is independent of $n_e$.
\subsection{Many-body expansion of the annihilation amplitude}
The many-body theory of positron annihilation $\gamma$-spectra was introduced in \cite{0953-4075-39-7-008}. 
In that work however, only the zeroth- and first-order correction to the annihilation vertex were discussed and the important effect of positron-atom correlations on the incoming positron wave function were neglected.  
In this section we give a concise overview of the theory. We present formulae for the full annihilation vertex and briefly discuss the treatment of the modified positron wave function. 

Taking the non-relativistic limit of the two-photon QED annihilation amplitude leads to the second 
quantized operator of two-photon annihilation $\hat{O}_a({\bf P})$ \cite{PhysRev.77.205,changlee,RevModPhys.28.308,0953-4075-39-7-008}
\begin{eqnarray}
\hat{O}_a({\bf P})=\int d^3r\; e^{-i{\bf P}.{\bf r}}{\hat \psi}({\bf r}){\hat \varphi}({\bf r}),
\end{eqnarray}
where ${\hat \psi}({\bf r})$ and ${\hat \varphi}({\bf r})$ are the electron and positron field annihilation operators respectively\,\footnote{
In this form of the operator the spin indices of the field annihilation operators have been suppressed, and summation over them is assumed.
This form can be used in systems with paired electron spins or when averaging over the positron spin. 
The modulus-squared amplitude is then multiplied by the spin-averaged QED factor $\pi r_0^2c$.
In general, one should use the spin-singlet combination of the annihilation operators, $\frac{1}{\sqrt 2}\left(\hat\psi({\bf r})\hat\psi({\bf r})-\hat\psi({\bf r})\hat\psi({\bf r})\right)$, together with the two-photon annihilation factor $4\pi r_0^2c$} . 
In this form, it is clear that non-relativistic annihilation occurs when the positions of the particles coincide.
To apply the many-body theory, a complete orthonormal set of single particle wave functions must be introduced. 
The expansion takes its simplest form in Hartree-Fock (HF) single-particle basis $|\mu\rangle$, 
in which the field operators of the electron and positron become 
\begin{eqnarray}
{\hat \psi}({\bf r})=\sum_{\mu} \psi_{\mu}({\bf r})\;\hat{a}_{\mu}\;\;;\;\; {\hat \varphi}({\bf r})=\sum_{\mu} \varphi_{\mu}({\bf r})\;\hat{b}_{\mu},
\end{eqnarray}
where $\psi_{\mu}({\bf r})$ and $\varphi_{\mu}({\bf r})$ are the HF wave functions (of central-field type),
and $\hat{a}_{\mu}
$ and $\hat{b}_{\mu}$ are the HF field annihilation operators, of the electron and positron.
In this basis the operator of annihilation becomes:
\begin{eqnarray}
\hat{O}_a({\bf P})=\sum_{\mu,\nu}\langle{{\bf P}}|\delta|\mu\nu\rangle\;\hat{a}_{\mu}\hat{b}_{\nu},
\end{eqnarray}
where we use the notation
\begin{eqnarray}\label{eqn:defpmunu}
\langle{{\bf P}}|\delta|\mu\nu\rangle \equiv \int e^{-i{\bf P.r}}\psi_{\mu}({\bf r})\varphi_{\nu}({\bf r})d^3r
\end{eqnarray}
The Hamiltonian of the system is 
\begin{eqnarray}
\hat{H}=\hat{H}_{\rm HF}+H_{\rm int},
\end{eqnarray}
where $\hat{H}_{\rm HF}$ is the Hartree-Fock Hamiltonian that satisfies $\hat{H}_{\rm HF}|\Phi_0^N\rangle=E_0^{\rm HF}|\Phi_0^N\rangle$, where $|\Phi_0^N\rangle=\prod_{i=1}^N a^{\dag}_i|0\rangle$ is the $N$-electron ground-state HF determinant wave function.
The many-body expansion of $A_{n\varepsilon}({\bf P})$ proceeds through the interaction Hamiltonian $H_{\rm int}\equiv\hat V_{\rm res}+\hat O_a({\bf P})$, where the residual interaction $\hat{V}_{\rm res}=\hat{V}_{e^-e^+}+\hat{V}_{e^-e^-}-\hat{U}^N_{\rm HF}$ is the sum of the standard two-body electron-positron and electron-electron Coulomb operators minus 
the one-body HF operator (for the frozen core of the $N$-electron system).

The resultant diagrammatic expansion is shown in \fig{fig:anndiagsgen} (see the figure caption for an explanation of the diagrams).
The construction of the diagrams follows standard rules.
First, note that the use of HF single-particle basis ensures that only pure Coulomb operators appear in the final diagrams, i.e., there are no \emph{explicit} contributions from $U_{\rm HF}$.  
Furthermore, since there is a single incident positron, only terms in the expansion involving a single $\hat{O}_a({\bf P})$ are non-zero \cite{DGG_thesis}.
Single-hole final-state diagrams at order $m=m_1+m_2$ in the Coulomb interaction (wavy line), where $m_1$ is the number of electron-positron Coulomb interactions and $m_2$ is the number of electron-electron Coulomb interactions, have $m_1$ positron propagators (single directed lines) and $m_1+2m_2$ electron propagators, in addition to a single positron and hole wave function (labelled $\varepsilon$ and $n$ respectively), and the two $\gamma$-rays (double-dashed lines).

\subsubsection{The annihilation vertex}
\begin{figure*}[!htb]
\includegraphics*[width=0.8\textwidth]{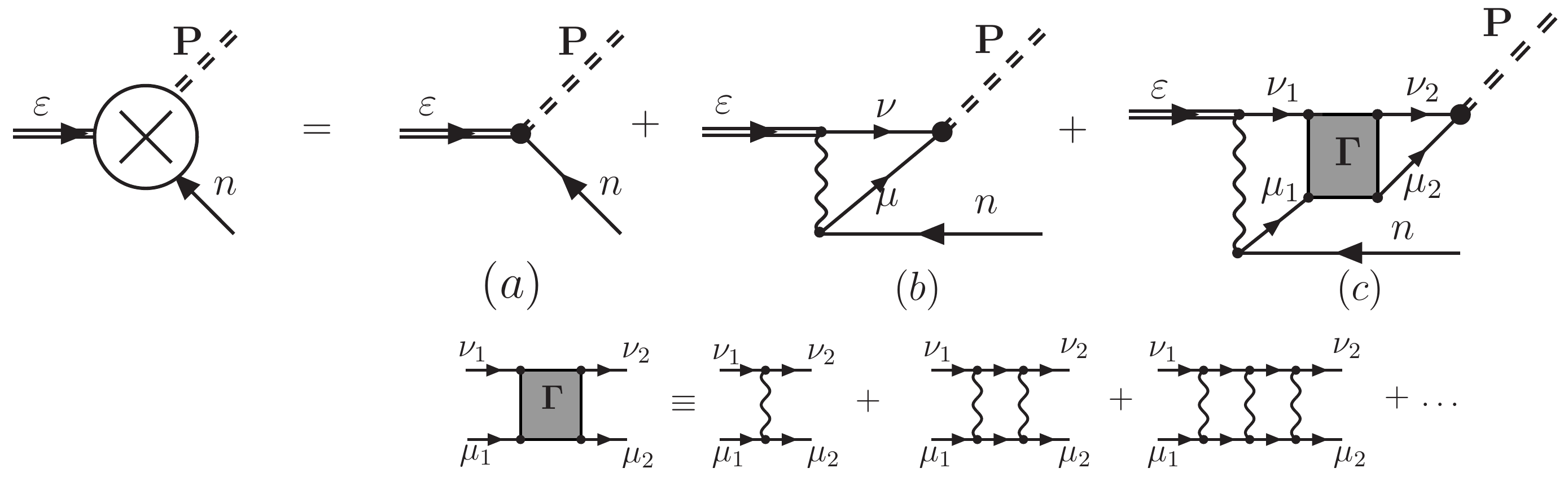}
\caption{Main vertex contributions to the annihilation amplitude $A_{n\varepsilon}({\bf P})$. 
Shown on the right hand side are the different orders of the annihilation vertex: (a) zeroth-order (IPA); (b) first-order correction, and; (c) `$\Gamma$-block' correction that contains the electron-positron ladder series.
\label{fig:anndiags}}
\end{figure*}

Figure \ref{fig:anndiags} shows the main contributions to the annihilation vertex: the zeroth-order (IPA) vertex, as well as the non-local first-order and the positron-electron ladder series, or `$\Gamma$-block', corrections.
From inspection of the diagrammatic expansion, one can write the exact annihilation amplitude in the abstract form
\begin{eqnarray}\label{eqn:annampgeneral}
\begin{split}
A_{n\varepsilon}({\bf P})~& =\int e^{-i{\bf P}\cdot{\bf r}}\left\{ \psi_{\varepsilon}({\bf r})\varphi_n({\bf r})\right.\\
&+ \left.\tilde\Delta_{\bf P}({\bf r};{\bf r}_1,{\bf r}_2)\psi_{\varepsilon}({\bf r}_1)\varphi_n({\bf r}_2) d{\bf r}_1d{\bf r}_2\right\} d{\bf r},
\end{split}
\end{eqnarray}
where the first term is the zeroth-order (IPA) vertex (\fig{fig:anndiags}a), and 
$\tilde\Delta_{\bf P}$ is the non-local annihilation kernel. 
In this form, the amplitude $A(\bf P)$ is clearly seen to be a Fourier transform of the correlated pair wave function (the term in the brace), whose modulus squared can be referred to as the electron-positron \emph{annihilation momentum density}. 
In the HF basis it takes the explicit form (using the notation of \eqn{eqn:defpmunu})
\begin{eqnarray}\label{eqn:annamp_details}
A_{n\varepsilon}({\bf P})&=&\langle{{\bf P}|\hat\bigotimes|n,\varepsilon}\rangle,
\end{eqnarray}
where the vertex operator
\begin{eqnarray}
\begin{split}
\hat\bigotimes&\equiv
\delta - \sum_{\mu,\nu}  \frac{\delta|\mu,\nu\rangle \langle\nu,\mu| V }{\varepsilon+\varepsilon_n-\varepsilon_{\mu}+\varepsilon_{\nu}}\\
&+\sum_{\mu_{i},\nu_{i}}  \frac{\delta|\mu_2,\nu_2\rangle \langle{\nu_2,\mu_2| \Gamma_{\varepsilon+\varepsilon_n} |\mu_1,\nu_1}\rangle  \langle{\nu_1,\mu_1| V }}{(\varepsilon+\varepsilon_n-\varepsilon_{\mu_2}-\varepsilon_{\nu_2})(\varepsilon+\varepsilon_n-\varepsilon_{\mu_1}-\varepsilon_{\nu_1})},
\end{split}
\end{eqnarray}
where the sum over $\mu$ and $\nu$ is over the excited electron and positron states.
Equation \ref{eqn:annamp_details} is written to emphasize the components common to all diagrams and is in accordance with the general diagrammatic form shown in Figs. \ref{fig:anndiagsgen} and \ref{fig:anndiags}: the external positron wave function (labelled $\varepsilon$), linked through the general vertex $\bigotimes$ to the annihilated electron (hole) wave function (labelled $n$) and the two-gamma rays ${\bf P}$. 
Note that the zeroth-order (IPA) result is just that of \eqn{eqn:defpmunu}, i.e., $A_{n\varepsilon}^{(0)}({\bf P})\equiv\langle {\bf P}|\delta|n\varepsilon\rangle$.
The $\Gamma$-block is determined from the linear integral equation, arrived at trivially by inspection of its diagrammatic form (shown in \fig{fig:anndiags})
\begin{eqnarray}
\langle \alpha\beta|\Gamma_E|\gamma\delta\rangle= \langle \alpha\beta|V|\gamma\delta\rangle + \sum_{\nu,\mu} \frac{\langle \beta|V|\mu\nu\rangle\langle \nu\mu|\Gamma_E|\gamma\delta\rangle}{E-\varepsilon_{\mu}-\varepsilon_{\nu}+i\eta}\nonumber,\\
\end{eqnarray}
where $\eta$ is a number close to zero. 
The successful computation of the $\Gamma$-block correction to the vertex was first performed in this way in \cite{PhysRevA.70.032720}, in which calculations of $Z_{\rm eff}$ for positron annihilation on hydrogen were reported. It was later incorporated in the calculations of the $\gamma$-spectra for annihilation on valence shells of the noble gases \cite{Dunlop_thesis}, and recently to annihilation on H-like ions \cite{DGG_hlike}. It can give rise to a significant enhancement above that of the first-order correction.

The Hartree-Fock wave functions involved in the matrix elements are of central field type $\varphi_{n\ell m}({\bf r})= \frac{1}{r}P_{n\ell}(r)Y_{\ell m}(\hat{\bf r})$.
The incident positron wave function is a continuum state of the form
\begin{eqnarray}
\psi_{{\bf k}}({\bf r})=\frac{\sqrt{4\pi}}{{r}}\sqrt{\frac{\pi}{k}}\sum_{\ell m}i^{\ell}e^{i\delta_{\ell}}
Y^\ast_{\ell m}(\hat{\bf k}) Y_{\ell m}(\hat{\bf r})
P_{\varepsilon \ell}(r)\nonumber,\\
\end{eqnarray} 
where $P(r)$ is normalized to a delta function of energy (in Rydbergs).
The positron radial wave function $P(r)$ is modified from the HF one by the positron-atom correlations, as explained below.

\subsubsection{Positron `Dyson' quasiparticle wave function}
The correlated incident positron \emph{quasiparticle} wave function $\psi_{\varepsilon}$ can be calculated 
from the integral form of the Dyson equation for the positron Green's function (see, e.g., \cite{abrikosov,fetterwalecka}): 
\begin{eqnarray}\label{eqn:dyson}
\left(H_0+\hat{\Sigma}_{\varepsilon}\right)\psi_{\varepsilon}({\bf r})=\varepsilon\psi_{\varepsilon}({\bf r}),
\end{eqnarray}
where $H_0$ is the unperturbed Hamiltonian of the $N+1$ particle system, which we take in this work to be the Hartree-Fock Hamiltonian, and $\hat{\Sigma}_{\varepsilon}$ is the non-local correlation potential. 
It is an integral operator that acts as $\hat{\Sigma}_{\varepsilon}\psi_{\varepsilon}({\bf r)}\equiv\int \Sigma_{\varepsilon}({\bf r},{\bf r}')\psi_{\varepsilon}({\bf r'})d{\bf r}'$, and is equivalent to the energy-dependent irreducible self-energy of the positron in the presence of the atom.
\begin{figure}[hbt!]
\includegraphics*[width=0.5\textwidth]{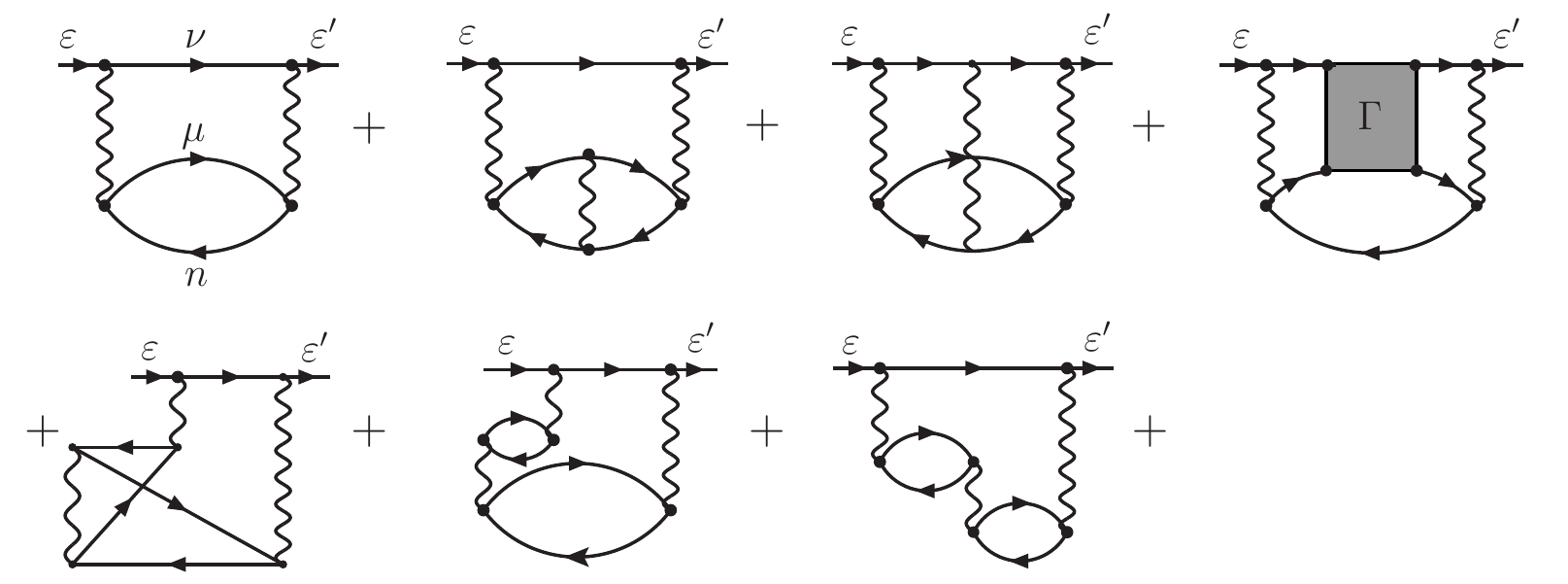}
\caption{The main contributions to the irreducible positron self-energy ${\Sigma}$, or more specifically, the matrix elements $\langle \varepsilon|\Sigma_E|\varepsilon'\rangle$. 
\label{fig:selfenergy}}
\end{figure}
The main contributions to $\Sigma$ are shown in \fig{fig:selfenergy}. 
At large positron-atom distances $\Sigma$ reduces to the local dipole polarization potential $\Sigma\approx-\frac{\alpha_d}{2r^4}\delta({\bf r-r}')$, where $\alpha_d$ is the dipole polarizability of the atom in the HF approximation
\begin{eqnarray}
\alpha_d=\frac{2}{3}\sum_{n,\mu}\frac{|\langle \mu|{\bf r}|n\rangle|^2}{\varepsilon_{\mu}-\varepsilon_n}.
\end{eqnarray}

Formally, the quasiparticle wave function (or `\emph{Dyson orbital}') $\psi_{\varepsilon}$ is given by
\begin{eqnarray}\label{eqn:quasiparticle}
\psi_{\varepsilon}({\bf r})=\int  \Psi^{*N}_0({\bf r}_1, \ldots, {\bf r}_N) \Psi^{N+1}_E({\bf r}_1, \ldots, {\bf r}_N; {\bf r}) d^N{\bf r},\nonumber\\
\end{eqnarray} 
i.e., it is the projection of  the fully-correlated $N+1$ particle wave function $\Psi^{N+1}$  (of energy $E$) onto the $N$-electron atomic ground state $\Psi_0^{N}$ of $N$ (of energy $E_0$), and $\varepsilon=E-E_0$.  
It can be calculated as outlined in Ref. \cite{PhysRevA.70.032720}, as
\begin{eqnarray}
\psi_{\varepsilon}({\bf r})=\varphi^{(0)}_{\varepsilon}({\bf r})+{P}\int \varphi_{\varepsilon'}({\bf r})\frac{\langle\epsilon'|\tilde\Sigma_{\varepsilon}|\varepsilon\rangle}{\varepsilon-\varepsilon'}d\varepsilon'.
\end{eqnarray}
where $\tilde\Sigma$ is the \emph{reducible} self-energy and where `$P$' means the principal value of the integral.
To normalise this wave function to a delta-function in energy (Rydbergs), it must be multiplied by the normalisation factor
\begin{eqnarray}
N(\varepsilon)=\left(1+\langle \varepsilon|\tilde{\Sigma}_{\varepsilon}|\varepsilon\rangle^2\right)^{-1/2}.
\end{eqnarray}
Overall, this method of determining the positron quasiparticle wave function is equivalent to solving the Schr\"{o}dinger equation with an optical potential \cite{PhysRevLett.3.96}.

\section{Numerical implementation}\label{sec:numerics}
The numerical procedure for evaluating the diagrams has been successfully demonstrated previously in Refs.~\cite{PhysRevA.70.032720} and \cite{0953-4075-39-7-008},
where it was used to calculate $Z_{\rm eff}$ for annihilation on atomic hydrogen, and the $\gamma$-spectra for annihilation on the valence shells of the noble gases.
The electrons involved in these studies have relatively small ionization energies, compared to the core-electrons that are of interest here.
More recently, the authors performed a study of positron annihilation on hydrogen-like ions \cite{DGG_hlike}. 
The ions considered there had electron ionization energies similar to those of typical core electrons of many-electron atoms. 
Using other, variational results as a benchmark, this study provided verification of the applicability and accuracy of the numerical method to annihilation with electrons of relatively large binding energies.
Full details of the numerical procedure can be found in the afforementioned works. 
Here, we aim to reduce repetition to the minimum compatible with the purpose of this section.

The main quantities to be calculated are the self-energy, the annihilation amplitude, and the $\gamma$-spectra $\bar w(\epsilon)$, whose angular reduced formulae are given in Appendix \ref{sec:appendix_angularreduction} and Ref. \cite{PhysRevA.70.032720}.
To evaluate the various many-body diagrams, one must first generate a set of single-particle Hartree-Fock basis states.
For the electron, this basis consists of the atomic ground-state wave functions as well as excited electrons in the field of the $N$-electron frozen core of the atom. 
For the positron, the static-field positron-atom potential is repulsive, and all states are continuum states in the field of the $N$-electron frozen core of the atom. 
To evaluate these, we use a mesh in momentum space of 201 points starting from threshold with a step size of $k=0.02$\au\

To facilitate the calculation of the diagrams, particularly those involving the ladder block, we introduce a B-spline basis \cite{splines,sapirstein_splines, 0034-4885-64-12-205,PhysRevA.70.032720,0953-4075-39-7-008,DGG_hlike}.
B-splines of order $k$ are piecewise polynomials of degree $k-1$ defined over a restricted domain (`box') that is divided into $n-k+1$ segments by a knot sequence of $n$ points: $r_i\in[0,R]$, where $n$ is the number of splines in the basis \cite{splines}.
In this work, we use an exponential knot sequence defined as $r_i=\rho_0(e^{\sigma i}-1)$, where $\rho_0=10^{-3}$\au\ and $\sigma$ is determined by the condition $r_{n-k+1}=R$, the box size. 
It allows for the accurate description of both the ground-state atomic wave functions, which can have many oscillations over the typical size of the atom and rapidly vanish outside the atom, as well as the continuum states that extend to larger distances up to the confinement (box) radius. 
The choice of the exponential knot sequence also ensures rapid convergence in the summations over intermediate states in the diagrams. The choice of spline basis parameters and box size is discussed below.

By expanding the single-particle HF orbitals $P_{n\ell}(r)=\sum_i c^{n\ell}_{i} B_i(r)$, where $c^{n\ell}_{i}$ are the expansion coefficients, and $B_i$ is the $i$-th spline of the basis, the Schr\"odinger equation can be reduced to the discrete eigenvalue problem
${\bf H}^{(\ell)}{\bf c_{\varepsilon}}=\varepsilon{\bf B}{\bf c_{\varepsilon}}$,
where $H^{(\ell)}_{ij}\equiv\int_0^R B_i(r)H_0^{(l)}B_j(r)dr$ and $B_{ij}\equiv\int_0^R B_i(r)B_j(r)dr$.
Note that to implement the boundary conditions $P(0)=P(R)=0$ the first and last splines are discarded in the expansions. 
The solutions of the equation for a given angular momentum $\ell$ are a set of $n-2$ eigenfunctions.
For the electron, the lowest energy states from this set correspond directly to the ground-state wave functions. 
The remaining electron states of the set describe excited electrons propagating in the frozen-core Hartree-Fock field of the $N$ ground-state electrons.
The solutions for the positron are continuum states that describe the positron propagating in the field of the HF frozen core of the $N$-electrons.

The choice of optimal spline basis parameters and box size depends on a number of competing considerations. 
These have been discussed in detail in Refs.~\cite{DGG_hlike} and \cite{PhysRevA.70.032720}.
For example, the vertex corrections to the annihilation amplitude involve small electron-positron distances and their accurate evaluation therefore requires an adequate minimum spatial resolution. 
Increased resolution can be obtained by either using a basis with a larger number of splines, or by reducing the box size, or both. For a given box size, however, increasing the number of splines in the basis reduces the density of states: a greater number of states must be included in the numerical calculations to achieve the same energy coverage and reach convergence. 
On the other hand, the positron-atom correlations involve are of longer-range and their accurate evaluation requires a large box size. 

\begin{figure}[!hp]
\includegraphics*[width=0.45\textwidth]{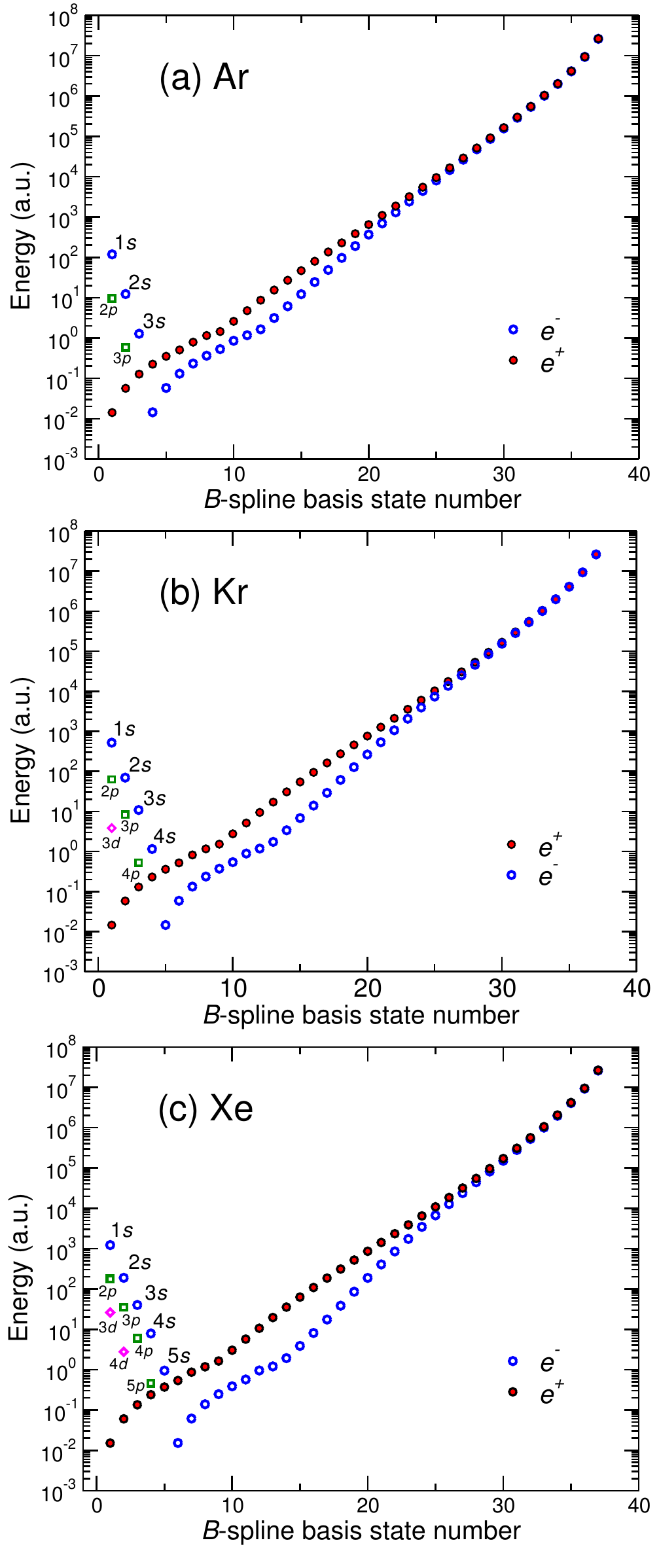}
\caption{\label{fig:basisenergies}Absolute values of the energy of B-spline basis states ($n=40$ splines of order $k=6$) with angular momentum $l=0$ for electron (empty circles) and positron (solid circles) for Ar, Kr and Xe. 
The $l=0$ electron ground states are labelled. 
For Ar, the energies of the $s$-states are 118.61, 12.32 and 1.28\au\ 
For Kr, the energies of the $s$-states are 520.17, 69.90, 10.85 and 1.15\au\
For Xe, the energies of the $s$-states are 1224.40, 189.34 and 40.18\au\
The ground $p$- and $d$-states are also shown as labelled squares and diamonds, respectively.
Spline states with $l>0$ have larger energies than those with $l=0$.} 
\end{figure}

The calculations were performed using a basis of $n=40$ splines of order $k=6$, and repeated using a basis of $n=60$ splines of order $k=9$.
For both spline bases, two different box sizes, $R=20$\au\ and $R=30$\au, were used.
We found that the $n=60$, $k=9$ core-subshell results suffered from poor convergence, and that the best overall results were obtained using the basis of $n=40$ splines of order $k=6$ with a box size of $R=20$\au\ 
The lowest 20 B-splines of the basis were included in the summations over intermediate states, corresponding to an energy coverage from zero to $\gtrsim 10^2$a.u., significantly greater than the energy of the valence and subvalence (core) ground-state wave functions that we are interested in (see \fig{fig:basisenergies}).
All results quoted henceforth have been calculated with these parameters.

When evaluating matrix elements involving the incident positron wave function $\varepsilon$, e.g., the self-energy matrix $\langle \varepsilon |\Sigma_{\varepsilon}|\varepsilon\rangle$, or the annihilation amplitude $A_{n\varepsilon}({\bf P})=\langle{\bf P}|\bigotimes|n,\varepsilon\rangle$, we use the true HF continuum states $|\varepsilon\rangle$.
This can be done by making further use of the spline completeness, introducing a resolution of the identity to re-write the matrix elements as, e.g., 
\begin{eqnarray}
\langle \varepsilon'|\Sigma_E|\varepsilon\rangle
&=&\sum_{\nu,\nu'}\langle \varepsilon'|f|\nu \rangle\langle \nu|f^{-1}\Sigma_Ef^{-1}|\nu'\rangle\langle \nu'|f|\varepsilon\rangle,\\
\langle {\bf P}|\delta|n,\varepsilon\rangle
&=&\sum_{\nu}\langle {\bf P}|\delta f^{-1}|\nu \rangle\langle \nu|f|n,\varepsilon\rangle.
\end{eqnarray}
The insertions of $f^{-1}f$, where $f\equiv(R-r)$, are made to nullify any numerical error arising due to the fact that the B-splines are zero at the boundary $R$, whereas the Hartree-Fock states are not.
In this way, the matrix elements of all quantities involving the true positron states can be evaluated as matrix elements involving individual spline basis states.

In the calculation of the positron self-energy (\fig{fig:selfenergy}) we include only those hole states $n$ in the diagrams that correspond to the two valence electrons, owing to the negligible contribution to the dipole polarizability of the atom $\alpha_d$ from the strongly-bound core electrons. 
Any small effect of omitting the core states from the self-energy diagrams will manifest only in the positron wave function, and should have a negligible effect on the vertex enhancement. 
In the diagrams of the annihilation amplitude the final hole state $n$ is, of course, taken to be any of the ground states, including the core subshells. 

The finite box size leads to a cut-off in the self-energy matrix element: it is evaluated only over the domain of the box. 
To compensate for this, we make use of the long-range asymptotic form of the correlation potential and add to the spline-based matrix element (calculated over the domain of the box) the contribution at $r>R$:
\begin{eqnarray}
\int_R^{\infty}P_{\varepsilon,l}(r)\left(-\frac{\alpha_d}{2r^4}\right)P_{\varepsilon'l}(r)dr,
\end{eqnarray}
where the asymptotic radial wave functions $P_{\varepsilon}$ are calculated with the HF phase shift as
\begin{eqnarray}\label{eqn:asympwf}
P_{\varepsilon l_{\varepsilon}}(r)\simeq r\sqrt{\frac{k}{\pi}}\left[j_{l_{\varepsilon}}(kr)\cos \delta_{l_{\varepsilon}}^{\rm HF} - n_{l_{\varepsilon}}(kr) \sin \delta^{\rm HF}_{l_{\varepsilon}}\right],~~
\end{eqnarray}
where $j_l$ and $n_l$ are the spherical Bessel and spherical Neuman functions, respectively.

For the calculation of the annihilation momentum distribution we used a mesh in the $\gamma$-ray momentum of 101 points from 0 to 9\,a.u. 
We also repeated the calculations with 101 points from 0 to 15\,a.u., although no discernible difference was found in the resulting $\gamma$-spectra in the range of Doppler-shifts of interest (0--9\,keV).

Finally, convergence with respect to the highest angular momentum of the intermediate excited states used in the calculation, $\ell_{\rm max}\to\infty$, was achieved using the asymptotic formula \cite{gfjl_extrapolation,0953-4075-39-7-008,DGG_hlike}
\begin{eqnarray}
\bar{w}_{n\varepsilon}^{[\infty]}(\epsilon)&=&w_{n\varepsilon}^{[\ell_{\rm max}]}(\epsilon)+{A}{(\ell_{\rm max}+1/2)^{-1}}\label{eqn:spectraextrapolation},
\end{eqnarray}
where, for a given $k$, $A$ is a constant that is determined from the results of the calculations performed using different $\ell_{\max}$. In this work all diagrams were calculated for $\ell_{\rm max}=7,8,9$ and $10$. 

\section{Results and Discussion}\label{sec:results}
In this section we present the results of the MBT calculations.
In \secref{sec:noblespectra} we discuss the effect of the correlations on the spectra of individual core and valence 
subshells of the noble gases Ar, Kr and Xe, and compare our calculated annihilation $\gamma$-spectra with the existing experimental spectra.
In addition, we calculate the total, and fractional contributions to, the normalized annihilation rate parameter $Z_{\rm eff}$ for each subshell. 
In \secref{sec:enhancement}, we use the results of the \abin\ many-body theory calculations to provide state-dependent enhancement factors for the core and valence electrons that quantify the vertex enhancement to the annihilation vertex. 
All the results presented have been calculated using an $s$-wave incident positron of thermal momentum ($k=0.04$\au).

\subsection{Annihilation $\gamma$-spectra of noble gases}\label{sec:noblespectra}
\subsubsection{Many-body calculated annihilation $\gamma$-spectra of valence and subvalence shells of Ar, Kr and Xe}
\begin{figure}[!hp]
\includegraphics*[width=0.472\textwidth]{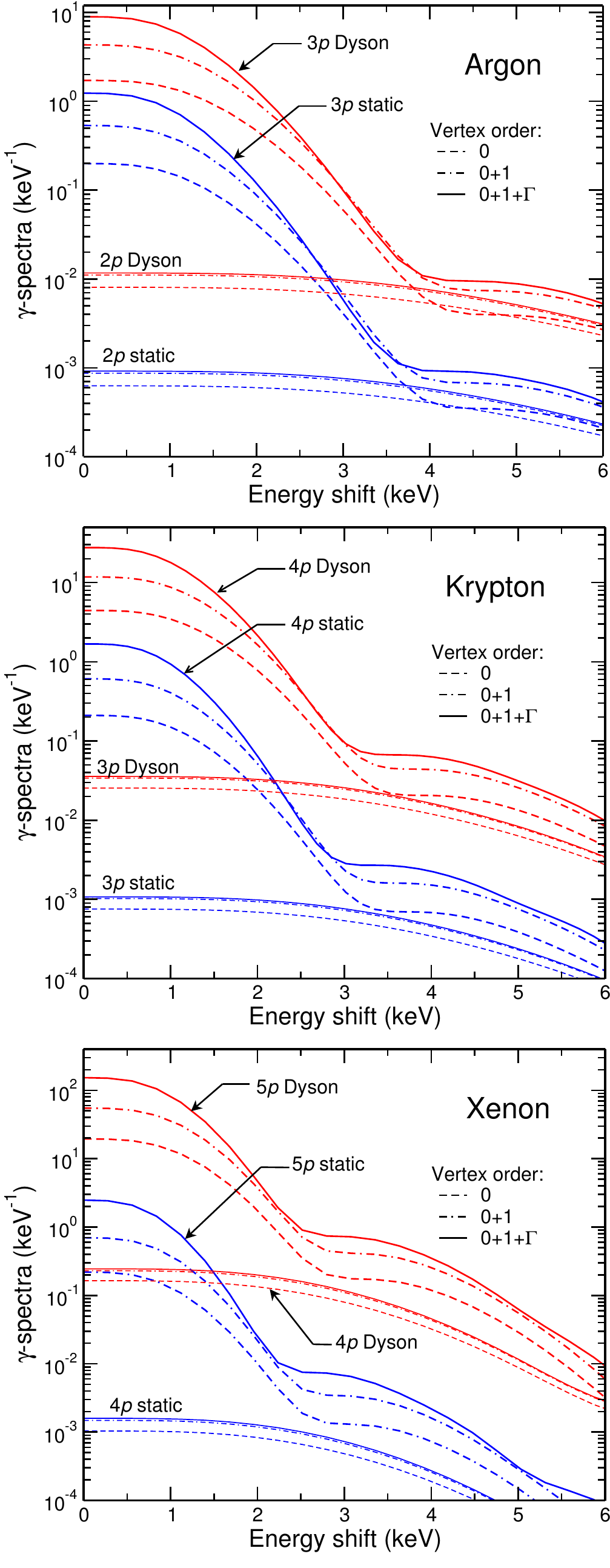}
\caption{$\gamma$-spectra for valence and subvalence $p$-subshells of Ar, Kr and Xe, 
calculated using HF (static) and Dyson positron wave function in different orders 
of the annihilation vertex.\label{fig:spectra_npnm1p}} 
\end{figure}

The effects of the correlations on the $\gamma$-spectra are shown in \fig{fig:spectra_npnm1p},
for the specific examples of the valence and subvalence $p$-subshells of Ar, Kr and Xe, respectively. 
For each subshell, the spectra calculated using both the HF (static) and Dyson incident positron in the three different approximations of the annihilation vertex (see \fig{fig:anndiags}) are shown. 
As the spectra are symmetric about zero energy shift, only positive Doppler energy shifts are shown.
Interpretation of the spectra is aided by \tab{table:zeffs}, which gives the values of $Z_{\rm eff}$ calculated as an integral of the spectra through \eqn{eqn:zeffspec}.

For each atom, for a given annihilation vertex order, the spectra calculated using the Dyson 
positron wave function are significantly greater in magnitude than those calculated using the HF one,
e.g., the spectrum of the valence $np$ subshell of Ar is increased above its Hartree-Fock result by $\approx$ 8 times; for Kr by $\approx 18$ times, and for Xe by as much as $\approx 70$ times. 
The reason for this \emph{wave function} enhancement is the introduction of an attractive potential through the strong positron-atom correlations. 
This potential supports low-lying positron virtual $s$ levels, which give rise to the characteristic resonant behaviour: $Z_{\rm eff}(k)\propto (\kappa^2+k^2)^{-1}$, where $\kappa=1/a$ is the reciprocal of the scattering length $a$, assumed here to be much larger than the radius at which annihilation occurs (typically the size of the atom) \cite{PhysLett.13.300,PhysScripta.46.248,dzuba_mbt_noblegas,PhysRevA.61.022720}. The relative strength of the wave function enhancement of Ar, Kr and Xe corresponds to the increase in their respective scattering lengths: for Ar $a\approx -4.4$\au, for Kr $a\approx -10.1$\au, and for Xe $a\approx -81$\au\ \cite{Ludlow_thesis}. 
Incidentally, this resonant behaviour means that $Z_{\rm eff}$ drops off dramatically as $k$ increases, and thus a proper comparison with the experimental $Z_{\rm eff}$ measured using thermalized positrons can only be made after performing a Maxwellian average of the calculated $Z_{\rm eff}(k)$ over $k$.  
Since it is the calculation of the $\gamma$-spectra that is the main interest here, this fact is not of utmost concern. 
Nevertheless, for comparison's sake the Maxwellian averaged $Z_{\rm eff}$ of the valence shells, calculated previously using MBT \cite{Ludlow_thesis}, are quoted in the table. 

For a given atom, deeper lying subshells experience a greater wave function enhancement. 
For example, for a given vertex order the Dyson spectrum for the Ar valence $3p$ subshell is $\approx$ 8 times larger than the HF spectrum, compared to the subvalence $(n-1)s$, for which it is $\approx$ 13 times larger. For Xe, the valence $4p$ subshell is $\approx$ 70 times larger than the HF spectrum, compared to the subvalence $(n-1)s$, for which it is $\approx$ 160 times larger.
This is a result of the fact that the positron must tunnel through a repulsive potential to reach the core electrons.
The additional attraction from the positron-atom correlations effectively lowers the HF repulsive potential: the wave function dies away more slowly in the tunnelling region, 
and consequently, the overlap of the positron wave function and the core electrons is increased.  
Comparatively, the overlap of the positron and valence electron wave functions is relatively unchanged.

\begin{table*}[htb!]
\caption{Calculated $Z_{\rm eff}$ (using \eqn{eqn:zeffspec}) and $\gamma$-spectra FWHM for annihilation on the valence $n$ and subvalence $(n-1)$ subshells of the noble gases.  Values are given for the different approxmations to the incident positron: HF orbital and Dyson quasiparticle wave function; and to the annihilation vertex: 0th, 0+1st and 0+1st+$\Gamma$ (\fig{fig:anndiags}). 
The values are for an $s$-wave incident positron of momenta $k=0.04$\au
\label{table:zeffs}}
\begin{ruledtabular}
\begin{tabular}{lc@{\hspace{2pt}}c@{\hspace{2pt}}c@{\hspace{2pt}}c@{\hspace{2pt}}c@{\hspace{2pt}}c@{\hspace{10pt}}cccccc}
Subshell		& \multicolumn{6}{c}{$Z_{\rm eff}$} &\multicolumn{6}{c}{$\gamma$-spectra FWHM (keV)}	\\
\cline{2-7}\cline{8-13}
			& \multicolumn{3}{c}{HF}	& \multicolumn{3}{c}{Dyson} & \multicolumn{3}{c}{HF}	& \multicolumn{3}{c}{Dyson}	\\
\cline{2-4}\cline{5-7}\cline{8-10}\cline{11-13}
			& 0		 	& 0+1 		& 0+1+$\Gamma$ 	& 0			 & 0+1 		& 0+1+$\Gamma$ & 0		 	& 0+1 		& 0+1+$\Gamma$ 	& 0	& 0+1 	& 0+1+$\Gamma$\\
\hline\\[-2ex]
			& \multicolumn{12}{c}{Argon}\\\\[-2.3ex]
$np$			& 0.606		& 1.530		& 3.144			& 5.651		& 13.43		& 24.97 		& 2.88	& 2.71 	& 2.40	& 3.12		& 2.95		& 2.63		\\
$ns$			& 0.136		& 0.260		& 0.343			& 1.387		& 2.576		& 3.320		& 1.87 	& 1.82	& 1.77	& 1.98 		& 1.93 		& 1.88		\\
$(n-1)d$		& $\cdots$	&$\cdots$		&$\cdots$			&$\cdots$		&$\cdots$		&$\cdots$ 	&$\cdots$	&$\cdots$	&$\cdots$	&$\cdots$		&$\cdots$		&$\cdots$		\\
$(n-1)p$		& 0.622[-2]	& 0.849[-2]	& 0.891[-2]		& 0.812[-1]	& 0.110		& 0.115 		& 9.42	& 9.28	& 9.24	& 9.57		& 9.43		& 9.39		\\
$(n-1)s$		& 0.235[-2]	& 0.307[-2]	& 0.319[-2]		& 0.308[-1]	& 0.400[-1]	& 0.414[-1]	& 5.19	& 5.20 	& 5.20	& 5.27		& 5.27		& 5.27		\\
Total			& 0.75		& 1.80 		& 3.50			& 7.15	 	& 16.16		& 28.44		& $\cdots$&$\cdots$&$\cdots$&$\cdots$&$\cdots$			& 2.52		\\
$\langle n~{\rm tot.}\rangle_{k}$ \footnote{Thermally averaged valence total from many-body theory, including additional diagrams \cite{Ludlow_thesis}.}	
			& $\cdots$	&$\cdots$ 	& $\cdots$		&$\cdots$	 	&$\cdots$		& 26.5		&$\cdots$&$\cdots$&$\cdots$&$\cdots$&$\cdots$	&$\cdots$				\\
Expt.\,\footnote{$Z_{\rm eff}$ total measured with thermalized positrons, from Ref.~\cite{coleman_noblegas_zeff}.}$^,$\footnote{Experimental $\gamma$-spectra FWHM from single-gaussian fit of Ref. \cite{PhysRevA.55.3586}. Note that a more accurate description of the experimental spectra is given by the two-gaussian fit of Ref. \cite{PhysRevA.55.3586}.}
			& $\cdots$	& $\cdots$ 	& $\cdots$		&$\cdots$ 	&$\cdots$		& 26.77$\pm$0.09	&$\cdots$&$\cdots$&$\cdots$&$\cdots$ 		&$\cdots$& 2.30\\\\[-2ex]
			& \multicolumn{12}{c}{Krypton}\\\\[-2.3ex]
$np$			& 0.565 		& 1.543 		& 3.742 			& 13.12 		& 16.48	 	& 68.67 		& 2.56	& 2.41	 & 2.12		& 2.81	& 2.67 	& 2.37		\\
$ns$			& 0.115		& 0.231		& 0.320			& 3.018		& 2.898		& 7.744		& 1.67	& 1.63	 & 1.57		& 1.78	& 1.75 	& 1.70		\\
$(n-1)d$		& 0.155[-1]	& 0.233[-1]	& 0.256[-1]		& 0.510		& 0.758		& 0.827		& 8.85	& 8.70	 & 8.64		& 9.09	& 8.95	& 8.88		\\
$(n-1)p$		& 0.599[-2]	& 0.805[-2]	& 0.843[-2]		& 0.205		& 0.271		& 0.283		& 7.65	& 7.58	 & 7.56		& 7.77 	& 7.70	& 7.68		\\
$(n-1)s$		& 0.172[-2]	& 0.221[-2]	& 0.229[-2]		& 0.591[-1]	& 0.754[-1]	& 0.779[-1]	& 4.35	& 4.35	 & 4.35		& 4.42	& 4.41	& 4.41		\\
Total			& 0.703		& 1.81		& 4.10			& 16.91	 	& 20.48		& 77.6		& $\cdots$&$\cdots$&$\cdots$&$\cdots$&$\cdots$		& 2.89		\\
$\langle n~{\rm tot.}\rangle_{k}$\,\footnotemark[1]
			& $\cdots$	&$\cdots$ 	& $\cdots$		&$\cdots$	 	&$\cdots$		& 66.4		&$\cdots$&$\cdots$&$\cdots$&$\cdots$&$\cdots$&$\cdots$\\
Expt.\,\footnotemark[2]$^,$\footnotemark[3]
			& $\cdots$	&$\cdots$ 	& $\cdots$		&$\cdots$	 	&$\cdots$		& 64.6$\pm$0.08	&$\cdots$&$\cdots$&$\cdots$&$\cdots$&$\cdots$				& 2.10\\\\[-2ex]
			& \multicolumn{12}{c}{Xenon}\\\\[-2.3ex]
$np$			& 0.510		& 1.535 		& 4.725 					& 50.806		& 136.67		& 334.01 		& 2.23	& 2.09	 & 1.81 		& 2.48	 & 2.37		& 2.07		\\
$ns$			& 0.104		& 0.228		& 0.351					& 11.98		& 24.64		& 35.77 		& 1.41	& 1.38	 & 1.33		& 1.55	 & 1.51 		& 1.46		\\
$(n-1)d$		& 0.229[-1]	& 0.387[-1]	& 0.450[-1]				& 3.525		& 5.671		& 6.602		& 6.73	& 6.65	 & 6.60		& 6.96	 & 6.89		& 6.84		\\
$(n-1)p$		& 0.628[-2]	& 0.892[-2]	& 0.952[-2]				& 1.010		& 1.403		& 1.490		& 5.79 	& 5.77	 & 5.76		& 5.91	 & 5.88		& 5.87		\\
$(n-1)s$		& 0.156[-2]	& 0.210[-2]	& 0.220[-2]				& 0.253		& 0.335		& 0.350 		& 3.35	& 3.36	 & 3.36		& 3.41	 & 3.41		& 3.41		\\
Total			& 0.645		& 1.81			& 5.13				& 67.57	 	& 168.0		& 378.2		& $\cdots$&$\cdots$&$\cdots$&$\cdots$&$\cdots$			& 2.01		\\
$\langle n~{\rm tot.}\rangle_{k}$\,\footnotemark[1]
			& $\cdots$	&$\cdots$ 	& $\cdots$				&$\cdots$	 	&$\cdots$		& 402		&$\cdots$&$\cdots$&$\cdots$&$\cdots$&$\cdots$&$\cdots$\\
Expt.\,\footnote{$Z_{\rm eff}$ total measured with thermalized positrons, from Ref.~\cite{xenon_zeff_trap}.}$^,$\footnotemark[3]
			& $\cdots$		&$\cdots$		&$\cdots$ 			&$\cdots$		&$\cdots$		& 401$\pm20$	&$\cdots$&$\cdots$&$\cdots$&$\cdots$&$\cdots$			&1.93
\end{tabular}
\end{ruledtabular}
\end{table*}
Concentrating now on the effect of the vertex corrections,  it is clear that the addition of the first and $\Gamma$-block vertex corrections enhances the spectra above the zeroth-order IPA result.
Fundamentally, this is due to the increase of the annihilation probability owing to the stronger attraction between the annihilating pair. This attraction is of particular importance, given the fact that non-relativistic annihilation occurs when the positions of the positron and electron coincide.
For both the core and valence shells, the addition of the first-order correction to the IPA vertex increases the magnitude of the spectra considerably: e.g., for the $4p$ valence and $3p$ subvalence subshells of Kr the spectra is increased at zero energy shift respectively by $\approx 2.5$ times from 4.36 to 11.44\,keV$^{-1}$, and by $\approx 1.3$ times from 0.025 to 0.034\,keV$^{-1}$.
For the valence shells, the addition of the higher order $\Gamma$-block correction causes a further significant enhancement: e.g., for the $4p$ subshell of Kr, it produces a further 2.3 times (from 11.44 to 27.6\,keV$^{-1}$) enhancement to the first-order result.
For the core however, the result of adding the $\Gamma$-block correction is much less dramatic, only raising the spectra slightly above that obtained with the first-order correction: e.g.,  for the $3p$ subshell of Kr, the $\Gamma$-block correction only increases the magnitude of the first-order result by a further 1.05 times (from 0.0340 to 0.0356\,keV$^{-1}$). 
A similar dominance of the first-order correction over the $\Gamma$-block correction was also seen for annihilation with the tightly-bound electrons of the H-like ions in \cite{DGG_hlike}.
Physically, the $\Gamma$-block describes virtual positronium formation by the tunnelling of an electron from the atom to the positron. 
This process is more difficult for the tightly bound spatially compact core than for the valence electrons.

From the figures it is also clear that for a given subshell, the enhancement from the vertex corrections is similar whether the HF or Dyson positron wave function is used. 
Physically, this is to be expected, as the vertex corrections are to a large extent independent of the positron-atom correlation.
We will discuss the vertex enhancement further in \secref{sec:enhancement}, where we evaluate enhancement factors that quantify the important vertex enhancement and show how they can be used in simple IPA type calculations to calculate accurate annihilation $\gamma$-spectra.

\begin{figure}[ht!!]
\includegraphics*[width=0.47\textwidth]{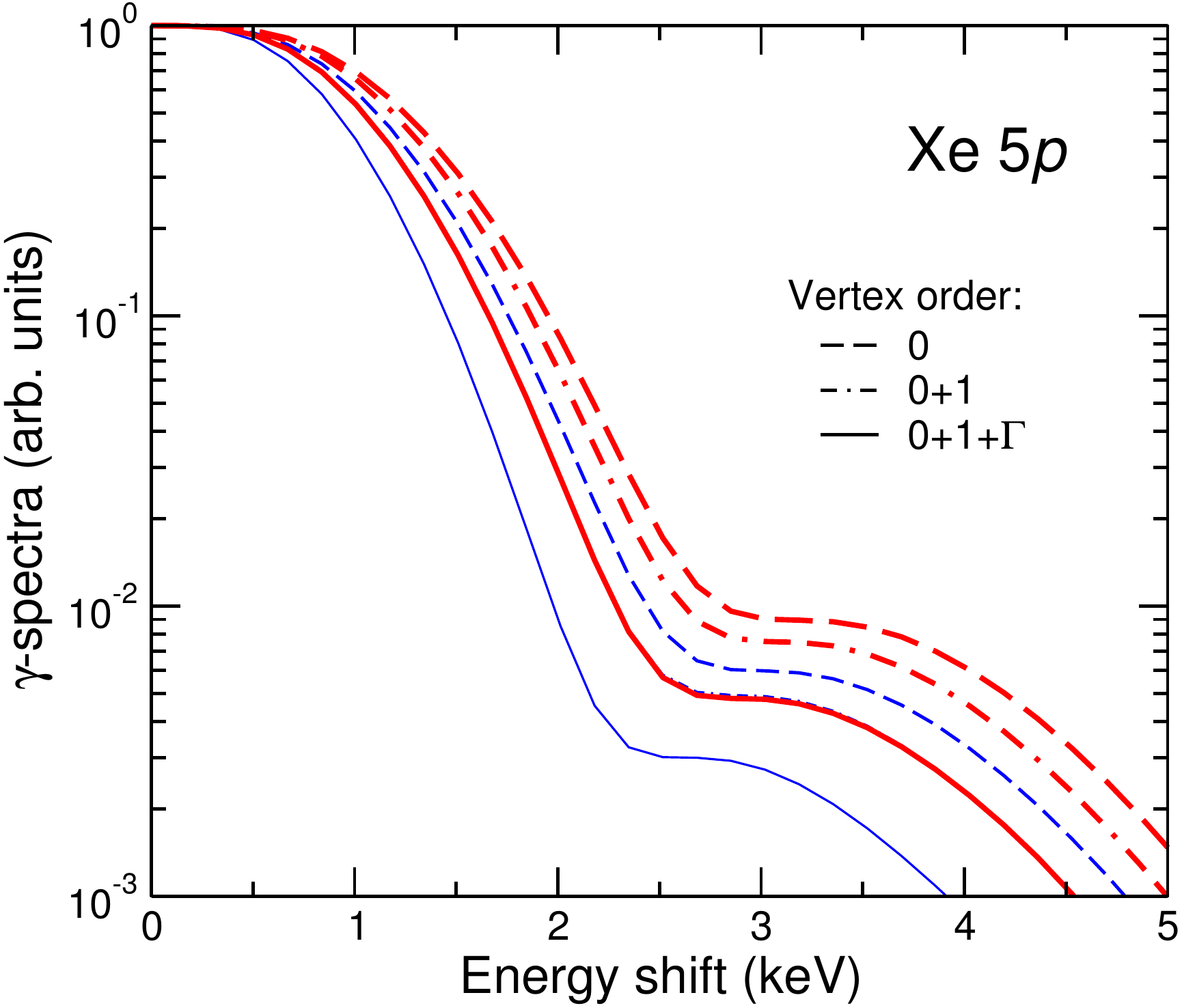}
\caption{The effect of electron-positron correlations on the $\gamma$-spectrum FWHM for the 5$p$
 subshell of Xe. Calculated using Dyson (heavy lines) and HF (light lines) positron in different orders of the annihilation vertex: zeroth (dashed); zeroth+ first (dash-dot line) and full (solid line). \label{fig:xe_spectra_fwhm}} 
\end{figure}
The correlations, both wave function and vertex type, evidently have a dramatic effect on the magnitude of the $\gamma$-spectra and the annihilation rate. 
Their effect on the shape of the spectrum is less so, although it is still significant.
\fig{fig:xe_spectra_fwhm} shows the spectrum of the Xe $5p$ subshell in the different approximations to the vertex and positron wave function. 
For ease of comparison, all curves  in the figure have been scaled to unity at zero energy shift.  
In addition, values of the full-width at half-maximum for all of the individual subshells calculated in the different approximations are given in \tab{table:zeffs}.
For the valence subshells, it is clear that increasing the self-energy corrections in the wave function for a given vertex order affects a broadening in the spectra: the attractive positron-atom correlations accelerate the positron towards the electron resulting in a relatively large momentum distribution, and hence a broader $\gamma$-spectrum. 
In contrast, increasing the vertex order affects a narrowing of the spectra: the first- and higher-order annihilation vertex corrections involve excited state (virtual) electrons that are relatively diffuse and therefore produce a relatively narrower momentum distribution, and hence a narrower $\gamma$-spectrum.
The degree of cancellation that occurs between this competing narrowing and broadening leads to a relatively small change in the full-width at half-maximum (FWHM) between the simplest zeroth-order-vertex HF-wave function result and the full MBT (0+1+$\Gamma$)-vertex Dyson-wave function result. 
This explains why one can obtain reasonable fractional core contributions through semi-empirical fitting of the zeroth-order HF-wave function calculated core and valence spectra to the total \cite{PhysRevLett.79.39}.
In contrast, the FWHM of the deeper lying subshells, e.g., the $(n-1)s$ subshell, does not change noticeably.
In summary, overall there is clearly some dependence of the FWHM on the approximation used, and reliable core contributions to the total spectra can therefore only be calculated by taking full account of the correlations. 

 \begin{figure*}[!ht]
\includegraphics*[width=1\textwidth]{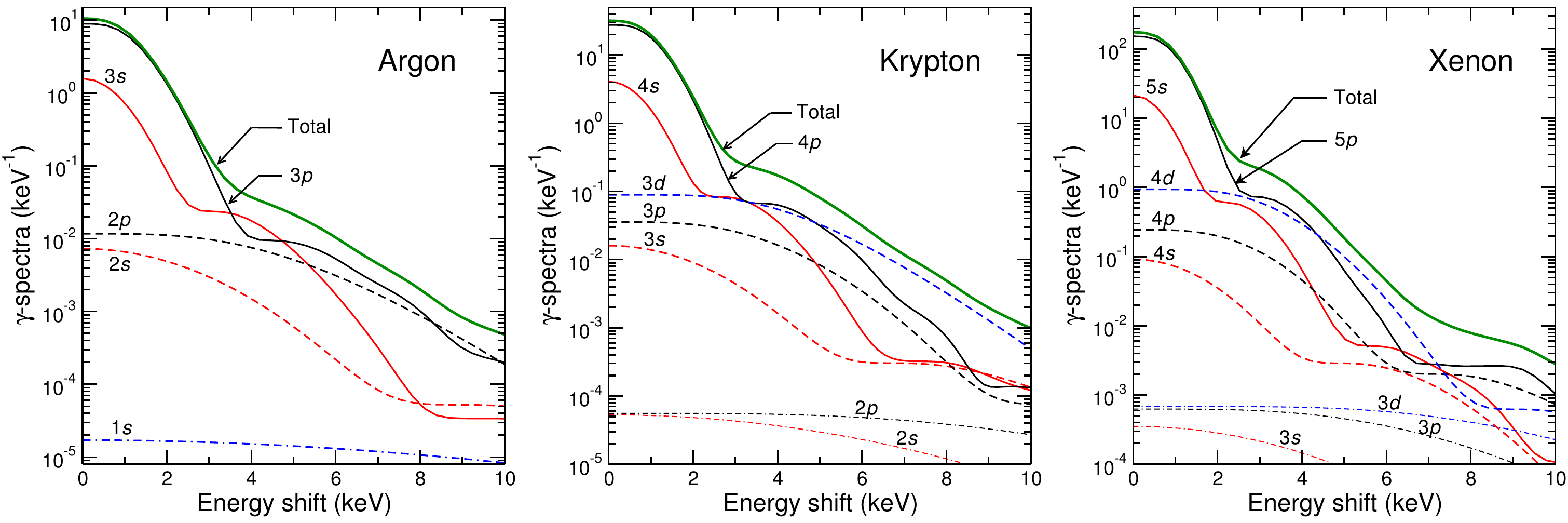}%
\caption{Annihilation $\gamma$-spectra for individual subshells of Ar, Kr and Xe,  
calculated with Dyson positron wave function and the full annihilation vertex. 
Only the valence $n$, sub-valence $(n-1)$ and $(n-2)$ subshells are shown for Kr and Xe as the more
inner shells contribute negligibly to the total. 
\label{fig:spectra}}
\end{figure*}

\fig{fig:spectra} shows the $\gamma$-spectra for the valence $n$, subvalence $(n-1)$ and $(n-2)$ shells for Ar, Kr and Xe, calculated using the full many-body theory (Dyson incident positron and full annihilation vertex) for $s$-wave incident positron. 
The corresponding tabulated values are given in Appendix \ref{sec:appendix_spectra}.
For each of the noble gases shown, the valence $np$-subshell is the main component of the total $\gamma$-spectra at small Doppler energy shifts ($\lesssim2$\,keV). 
This subshell is the most diffuse of the ground states.
By considerations of the uncertainty principle, it therefore contributes to the spectrum mainly at low momenta (small energy shifts). 
Similarly, at small energy shifts, the magnitude of the valence shells is much greater than that of the subvalence or deeper lying shells, e.g., the valence shell of Kr is more than two orders of magnitude greater than that of the subvalence shell, which is a further three orders of magnitude greater than the next most inner shell. 
The need for the positron to overcome an increasing nuclear repulsion as it penetrates closer to the core causes this reduction in the probability of annihilation. 

The characteristic shapes of the individual subshell spectra are governed by the properties of the annihilated electrons. 
As the angular momentum of the annihilated electron increases, the spectra become flatter at small energy shifts. Indeed, for an $s$-wave positron it is easy to show from the first-order result (of the form of \eqn{eqn:defpmunu}) and the small argument limit of the Bessel functions that the spectra behave as $w(0)-w(\varepsilon)\propto|\varepsilon|^{2\ell_n+2}$ for energy shifts $|\varepsilon|\ll 1$. 
Other features include the multiple `bell' shapes, or `shoulders', evident in the individual spectra. 
Looking at the subvalence $3s$ subshell of Ar in particular, three distinct bell shapes are visible: the first in the region 0--2\,keV, the second, with a much larger width, in the region 2--8\,keV, and the third in the region $>8$\,keV.
The number of distinct bell shapes in a given spectra is determined by the number of non-zero regions of energy density in the electron's radial wave function, $n-\ell$. 
They can be understood by appealing to the uncertainty principle. 
For example, the $3s$ radial wave function of Ar, which has 2 nodes, has three regions of non-zero wave function density. 
wave function density at small radii manifests in the high-momentum (energy) regions of the spectra. 
The regions of non zero wave function density at larger $r$ will contribute to the small momentum (energy regions) of the spectra. 
A further feature is that at large Doppler shifts the shape of the spectrum of the valence $np$-subshell, e.g., for Ar, the spectrum of the $3p$ valence subshell follows that of the subvalence $2p$ subshell; the requirement of orthogonality forces the $3p$ and $2p$ wave functions to be similar at small radii. 
Accordingly, their spectra are similar are large energy shifts (momenta).

{
\squeezetable
\begin{table}[tb!]
\caption{Calculated fractional contribution to annihilation rate for noble gases (thermal incident positron).\label{table:annfracs}} 
 \begin{ruledtabular}
 \begin{tabular}{lccccccc}
Subshell		& \multicolumn{2}{c}{Argon}	&  \multicolumn{2}{c}{Krypton}	 	& \multicolumn{2}{c}{Xenon}	\\
\cline{2-3}\cline{4-5}\cline{6-7}
			&HF-IPA	 	& MBT		&HF-IPA	 	& MBT		&HF-IPA		 & MBT			\\
\hline
$np$ 		& 0.8075 		& 0.8778		& 0.8032		& 0.8848		& 0.7907		& 0.8831			\\
$ns$			& 0.1810 		& 0.1167		& 0.1640		& 0.998[-1]	& 0.1616		& 0.946[-1]		\\				
$(n-1)d$		& $\cdots$	&$\cdots$ 	& 0.218[-1]	& 0.107[-1]	& 0.355[-1]	& 0.175[-1]		\\				
$(n-1)p$		& 0.830[-2]	& 0.405[-2]	& 0.852[-2]	& 0.365[-2]	& 0.973[-2]	& 0.394[-2]		\\				
$(n-1)s$		& 0.314[-2]	& 0.146[-2]	& 0.244[-2]	& 0.100[-2]	& 0.241[-2]	& 0.926[-3]		\\				
Total			& 1.000 		& 1.000		& 1.000		& 1.000		& 1.000		& 1.000 			\\
\hline				
$(n-1)$ tot.	&0.011		& 0.0055 		&0.0328		& 0.0154		& 0.0476	& 0.0224	\\	
$(n-1)$ tot.\footnote{Calculated in Ref. \cite{PhysRevLett.79.39} using semi-empirical fit.}	&\multicolumn{1}{c}{$<$0.002}&	$\cdots$	&\multicolumn{1}{c}{0.013}&$\cdots$	 	&\multicolumn{1}{c}{0.024} &$\cdots$\\		 \end{tabular}
 \end{ruledtabular}
 \end{table}
 }

\tab{table:annfracs} shows the contribution of individual subshells to the total spectrum as a 
fraction of unity. Values are given for both the simplest calculation that uses a HF incident positron with the IPA vertex, and for the full MBT 
calculation (full vertex with Dyson incident positron wave function).
For all of the noble gases, the inclusion of the electron positron correlations result in an increased fraction of annihilation on the valence shells, compared to the core. 
In fact, the HF-IPA calculation overestimates the total core contribution `$(n-1)$ tot.'~by a factor of $\sim2$.  
This is a result of the varying strength of the state dependent enhancement, which is stronger for the valence than for the inner shells.  
Also shown in the table are the results of the semi-empirical fitting of \cite{PhysRevLett.79.39},
in which the fractions were calculated by adjusting the ratio of HF-IPA inner to valence shell spectra in the total. 
For Kr and Xe this simplistic method gives fractional contributions that are of similar order to the parameterless MBT results, 
owing to the robustness of the shape of the spectra to the effect of correlations as discussed above (see \fig{fig:xe_spectra_fwhm}). This method cannot however, produce individual spectra of accurate magnitude like those of \fig{fig:spectra}. Such spectra are required to calculate state-dependent $\gamma$-spectra enhancement factors.

\subsubsection{Comparison with experimental $\gamma$-spectra: the core-electron contribution to the spectra}
\begin{figure}[!t]
\includegraphics*[width=0.48\textwidth]{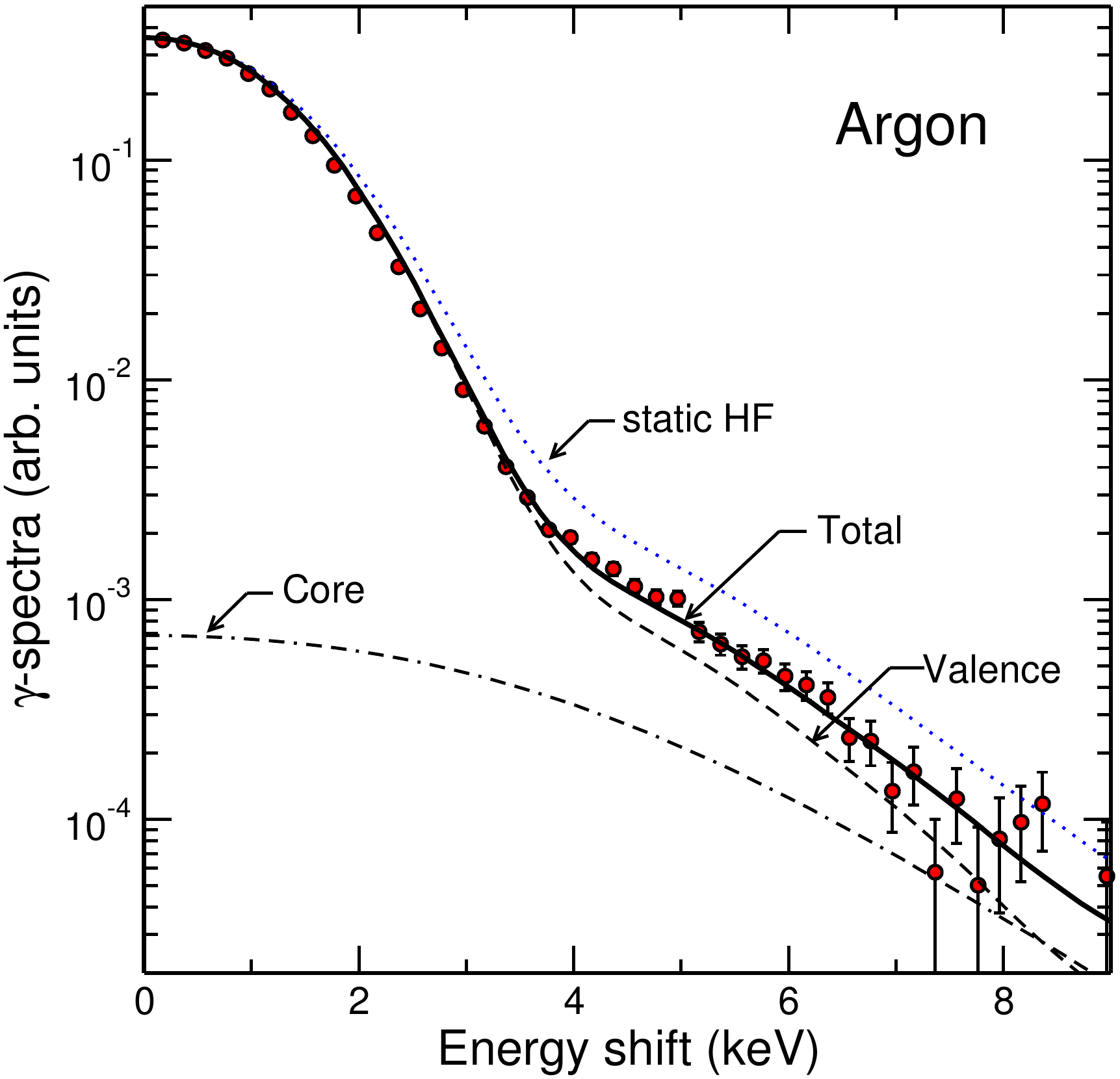}
\caption{Annihilation $\gamma$-spectra of Argon. MBT calculated inner- (chain curve) and 
valence (dashed curve) shell contributions to the total (solid) are compared with experiment 
(circles) \cite{PhysRevLett.79.39}. Also shown is the result of an IPA (zeroth-order vertex) HF calculation (dots). 
All calculated spectra have been scaled to experiment at zero energy shift. \label{fig:expspectra_ar}}
\end{figure}
\begin{figure}[!h]
\includegraphics*[width=0.48\textwidth]{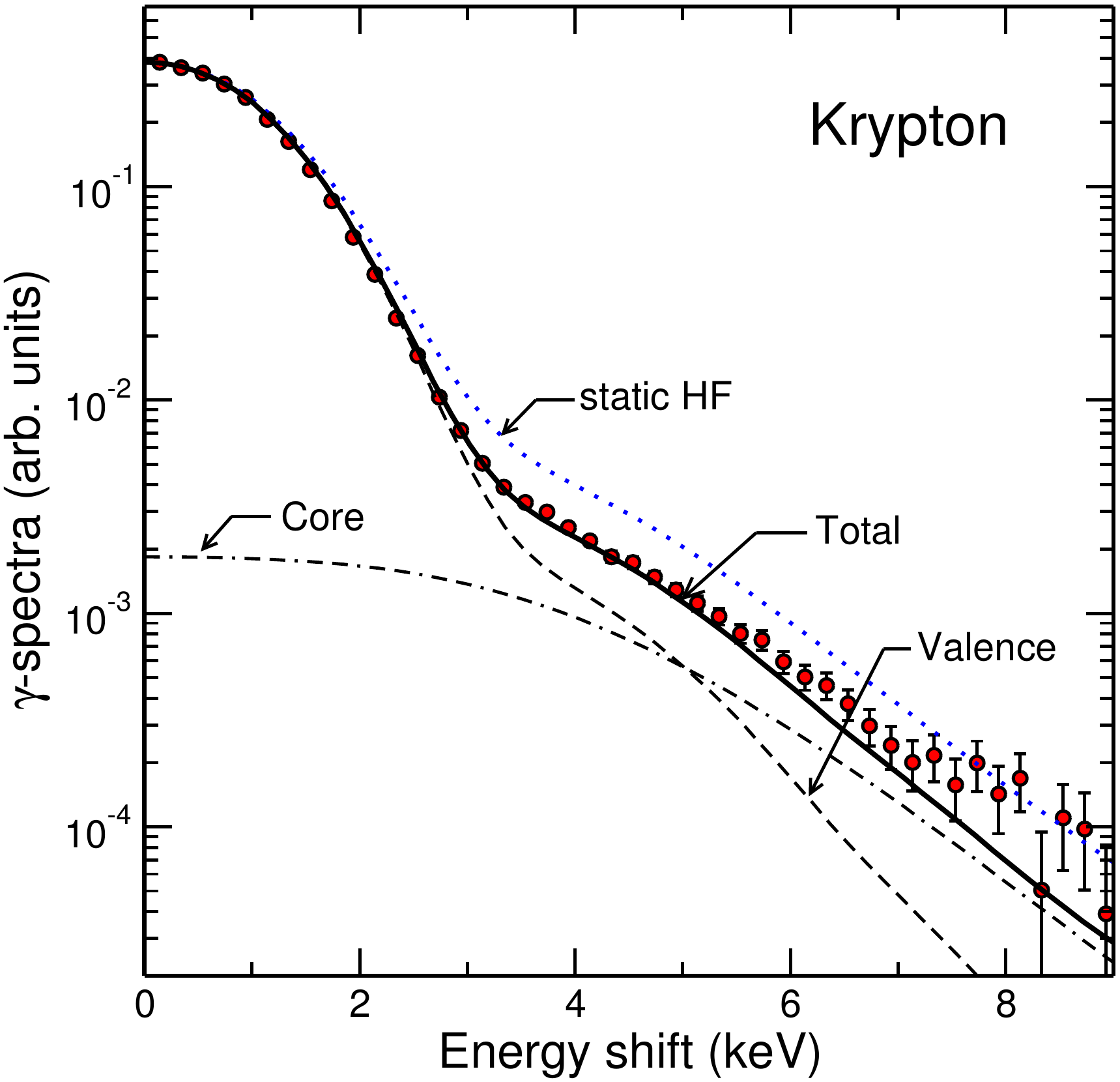}
\caption{Annihilation $\gamma$-spectra of Krypton (legend as described in caption 
of \fig{fig:expspectra_ar}).\label{fig:expspectra_kr}}
\end{figure}
\begin{figure}[!h]
\includegraphics*[width=0.48\textwidth]{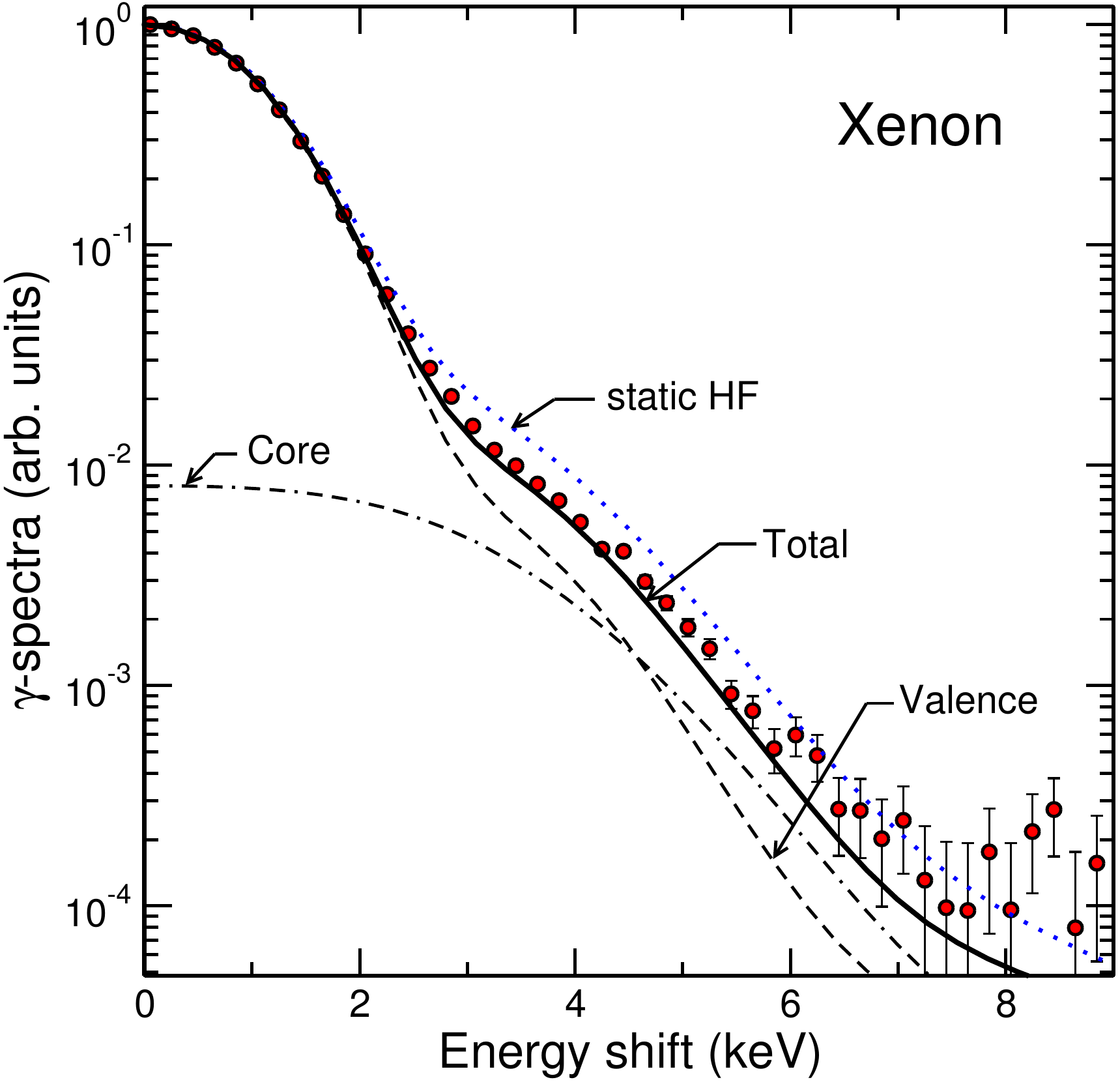}
\caption{Annihilation $\gamma$-spectra of Xenon (legend as described in caption 
of \fig{fig:expspectra_ar}).\label{fig:expspectra_xe}}
\end{figure}
\fig{fig:expspectra_ar}, \ref{fig:expspectra_kr} and \ref{fig:expspectra_xe} show the calculated spectra for Ar, Kr and Xe, after convolving with the experimental detector resolution function \cite{PhysRevLett.79.39}
\begin{eqnarray}\label{eqn:dres}
f(\epsilon)=N\exp\left({\frac{\epsilon}{a\Delta E_{\rm det}}}\right)^2,
\end{eqnarray}
where $a\equiv1/(4\ln2)^{1/2}$, $N$ is a normalization constant and $\Delta E_{\rm det}=1.16$\,keV is the width of the detector response,
and scaling to the experimental spectra at zero energy shift. 
Specifically, the figures show the individual contributions of the core and valence shells to the total spectra, calculated with the full MBT, the experimental results with associated errors, and the result of a simple HF-IPA calculation.
In the experimental results, the error bars represent the expected statistical 
variation in spectral amplitude due to counting statistics. 
Only the positive energy shifts of the symmetric $\gamma$-ray line is shown, 
since this region has greater statistical significance due to the reduced Compton scattering component \cite{PhysRevLett.79.39}.
In each figure the MBT calculated `valence' spectra are the sum of the spectra of the $ns$ and $np$ subshells. 
As was seen in \fig{fig:spectra},  the contribution of the $(n-2)$ and more inner shells are negligible compared to the subvalence shell.
In Figs.\,\ref{fig:expspectra_ar}, \ref{fig:expspectra_kr} and \ref{fig:expspectra_xe} the `inner' spectra are therefore calculated as just the sum of the spectra for the individual $(n-1)$ subvalence subshells: 
$(n-1)s+(n-1)p+(n-1)d$, or just $2s+2p$ for Ar. 
The relatively diffuse valence electrons, which have a narrow momentum distribution, dominate the spectrum at small energy shifts. 
This is clear from the figures, which show that the spectrum calculated from the valence 
shells alone agrees well with experiment at Doppler energy shifts $\lesssim2$\,keV, 
but is insufficient at larger energy shifts. 
In contrast, the core electrons are spatially compact, and therefore have a broader momentum distribution compared with the diffuse valence electrons. 
They contribute markedly to the high-energy `wings' of the spectrum, and their importance in bringing the total inline with experiment is evident for all three noble gases shown. 
For Ar, the inclusion of the core brings theory into excellent agreement with experiment over the full range of energy shifts.
For Kr the valence shell contribution alone severely underestimates the total spectrum, and the core contribution is of even greater importance than in Ar, dominating over the valence at energies $\gtrsim5$\,keV. 
Again the overall agreement of the spectra with experiment is excellent, although at higher energies the theoretical total slightly underestimates the experimental spectrum. 
In Xe, the valence shell again severely underestimates the total spectrum.
The core contribution dominates over the valence at energies $\gtrsim4.5$\,keV, and is again seen to be vital in achieving agreement with experiment. 
Similar to the case of Kr, however, the theoretical spectrum slightly underestimates the experimental one at large energy shifts, and is `steeper' than experiment. 
Since the theoretical results have been scaled to experiment, this suggests that the ratio of small to large momentum components is overestimated. 

One possible reason for the slight discrepancy is the neglect of relativistic effects on the electron wave functions. 
Indeed, Cook \emph{et al.} \cite{PhysRevLett.52.1116} have shown that the momentum distributions of the valence Xe subshells show clear and direct relativistic effects manifestations of relativistic effects.
In general, the relativistic core orbitals will be more spatially compact than their non-relativistic counterparts and will have a correspondingly larger momentum density\,\footnote{The valence shells also experience an less dominant indirect relativistic effect resulting from an increased nuclear screening},\cite{relqm}. 
Compared to the non-relativistic spectra, the high-energy regions of relativistic spectra should therefore be increased w.r.t.~the low-energy regions, corresponding to an overall flattening.
Such a flattening was evident in investigations of the Hartree-Fock and relativistic Dirac-Hartree-Fock (DHF) Compton profiles of the noble gases \cite{PhysRevA.75.022504}. 
However, in that study the HF and DHF profiles were found to be similar at the momenta corresponding to the high-energy `wings' of the Doppler spectrum and suggest that relativistic effects are not significant.
In addition, since relativistic effects on the spectra scale as $Z^2$ \cite{relqm}, one would expect to see a larger discrepancy in the spectra of Xe (Z=54) than of Kr (Z=36).
In contrast, we observe as similar discrepancy in both. 
A conclusive statement would require a full relativistic many-body calculation of the annihilation spectra.

One further possible source of the discrepancy arises from the fact that, in order to be completely accurate, one should perform a Maxwellian average of the magnitude of the raw $\gamma$-spectra over the incident positron momentum $k$, i.e., one should calculate the thermal averaged spectra given by
\begin{eqnarray}
\langle\bar{w}(\epsilon)\rangle_k=(\beta/2\pi)^{3/2}\int \bar{w}_k(\epsilon) e^{-{\frac{1}{2}{\beta k^2}}}4\pi k^2dk,
\end{eqnarray}
where $\beta\equiv 1/k_BT$ and $k_B$ is the Boltzman constant. 
This is likely to be broader than the spectra calculated at $k=0.04$\au\\\

Regardless, the discrepancy between the theoretical total and the experiment at the high energy shifts in Kr and Xe is small, and despite it the MBT results show unequivocally that: (1) the valence shells alone are insufficient to describe the total $\gamma$-spectra of Ar, Kr and Xe, and that; (2) there is a significant contribution from the relatively broad core spectrum to the total, and in addition, that this contribution is crucial to obtain agreement with the experimental spectra.

\subsection{Vertex enhancement factors for annihilation $\gamma$-spectra}\label{sec:enhancement}
It should by now be clear that the correlation corrections to the annihilation amplitude must be incorporated in an accurate calculation of the annihilation $\gamma$-spectra of the valence and core subshells. 
The full many-body theory calculation is, however, non-trivial, and its application numerically intensive. 
To interpret the results of, e.g., PAES experiments, one needs some way of predicting accurate core and valence annihilation spectra. 
Here, we appeal to the predilection of the literature and use the many-body theory to calculate \emph{enhancement factors}.
These factors quantify the enhancement that results from the non-local corrections to the annihilation vertex. 
Such factors are common in positron-atom, positronium \cite{PhysRevA.72.062707,PhysRevA.70.032720}, positron-charged-ion \cite{PhysRevB.20.883, PhysRevA.69.052702} and positron-condensed-matter studies \cite{PhysRevB.34.3820, RevModPhys.66.841, PhysRevB.51.4176,Barbiellini1997283,PhysRevB.56.7136,PhysRevB.54.2397,PhysRevB.73.035103,PhysRevB.58.10475,PhysRevB.57.12219}. 
They are, however, usually calculated in some phenomenological way.
They can be determined from an \abin\ viewpoint from many-body theory, simply by calculating the required quantities in different approximations to the vertex. 
and can be incorporated in simple IPA calculations to obtain accurate annihilation spectra. 

In general, one can define the momentum-dependent enhancement factor by approximating the non-local annihilation amplitude by the local, modified IPA form (\emph{cf.} Eqns. (\ref{eqn:annampgeneral}) and (\ref{eqn:annamp_details}))
\begin{eqnarray}\label{eqn:defenhance}
A_{n\varepsilon}({\bf P})\approx \int e^{-i{\bf P}\cdot{\bf r}} \psi_{\varepsilon}({\bf r})\psi_n({\bf r}) \sqrt{\bar\gamma_{n\varepsilon}({\bf r, P})}d{\bf r},~
\end{eqnarray}
where $\sqrt{\bar\gamma_{n\varepsilon}}$ is the state-dependent enhancement factor, usually interpreted as modifying the electron density at the positron (the square root is for later convenience).
The calculation of $\bar\gamma$ from many-body theory necessitates the reduction of our non-local annihilation amplitude to the IPA form of \eqn{eqn:defenhance}.
Formally, using \eqn{eqn:annamp_details} and \eqn{eqn:defenhance}, the true enhancement factor is given by
\begin{eqnarray}\label{eqn:bargammapformal}
\sqrt{\bar\gamma_{n\varepsilon}({\bf r, P})}\equiv 1+\frac{\iint \tilde\Delta_{\bf P}({\bf r};{\bf r}_1{\bf r}_2)\psi_{\varepsilon}({\bf r}_1)\varphi_n({\bf r}_2) d{\bf r}_1d{\bf r}_2}{\psi_{\varepsilon}({\bf r})\varphi_n({\bf r})}.\nonumber\\
\end{eqnarray}
The presence of nodes in the wave functions of the denominator, however, render this quantity of limited use and we must opt for a more pragmatic approach.

The enhancement is in general dependent on the two-$\gamma$ momentum ${\bf P}$ \cite{DGG_hlike}.
However, we will see that to all intensive purposes this momentum dependence does not carry through to the annihilation $\gamma$-spectra, at least for the core levels. 
It is therefore instructive to define a more simplistic state-dependent enhancement factor as the ratio of the calculated true partial annihilation rate (evaluated through \eqn{eqn:zeffspec}) to that calculated using the zeroth-order (IPA) vetex;
\begin{eqnarray}
\bar\gamma_{n}\equiv\frac{Z^{(0+1+\Gamma)}_{{\rm eff}, n}}{Z^{(0)}_{{\rm eff}, n}},
\end{eqnarray}
where the superscript denotes the vertex order (see \fig{fig:anndiags}) and $n$ labels the electron subshell. 
This parameterization is commonly used to predict true annihilation rates and $\gamma$-spectra (see, e.g., \cite{PhysRevB.51.4176,Barbiellini1997283,PhysRevB.56.7136,PhysRevB.54.2397,PhysRevB.73.035103,PhysRevB.58.10475}).
The true spectra for annihilation on a given subshell is then to be approximated by
\begin{eqnarray}\label{eqn:spectra_reconstructed}
\bar w_n(\epsilon)\approx \bar\gamma_n w^{(0)}_{n}(\epsilon).
\end{eqnarray}
This procedure therefore assumes the vertex enhancement is independent of the Doppler-energy shift.
\begin{table}[htb!]
\caption{Many-body theory calculated vertex enhancement factors $\bar{\gamma}_n(k_{\rm Th})$ for noble gas subshells, calculated for $s$-wave thermal ($k=0.04$\,\au) Hartree-Fock (HF) and Dyson (Dy) incident positron.\label{table:bargamma}}
\begin{ruledtabular}
\begin{tabular}{lccccccccc}
				& \multicolumn{2}{c}{Argon}	&\multicolumn{2}{c}{Krypton} &\multicolumn{2}{c}{Xenon}\\
\cline{2-3}\cline{4-5}\cline{6-7}
Subshell			& HF	&Dy				&HF&Dy				&HF&Dy	\\
\hline						
$np$				& 5.19 & 4.42			& 6.63 & 5.23			& 9.26 & 6.57	\\
$ns$				& 2.51 & 2.39			& 2.77 & 2.57			& 3.36 & 2.99	\\
$(n-1)d$			& $\cdots$ & $\cdots$	& 1.67 & 1.62			& 1.96 & 1.87	\\
$(n-1)p$			& 1.43 & 1.42			& 1.41 & 1.38			& 1.52 & 1.48	\\
$(n-1)s$			& 1.35 & 1.34			& 1.34 & 1.32			& 1.42 & 1.39	\\
\hline
$n$ tot. 			& 4.70	& 4.02		& 5.97	& 4.73		& 8.27 & 5.89 	\\
$(n-1)$ tot. 		& 1.41	& 1.40		& 1.56	& 1.53		& 1.85 & 1.76	\\
\end{tabular}
\end{ruledtabular}
\end{table}
\tab{table:bargamma} shows the values of $\bar\gamma$ for the valence $n$ and subvalence $(n-1)$ shells of Ar, Kr and Xe, calculated from the MBT using both the HF and Dyson incident positron.
For the valence electrons, the vertex enhancement is dramatic, e.g., it produces a five-times increase in the spectral magnitude of the diffuse $p$-subshell of Ar, and a nine-times increase for that of Xe (for the HF positron).
\begin{figure}[t!]
\includegraphics[width=0.48\textwidth]{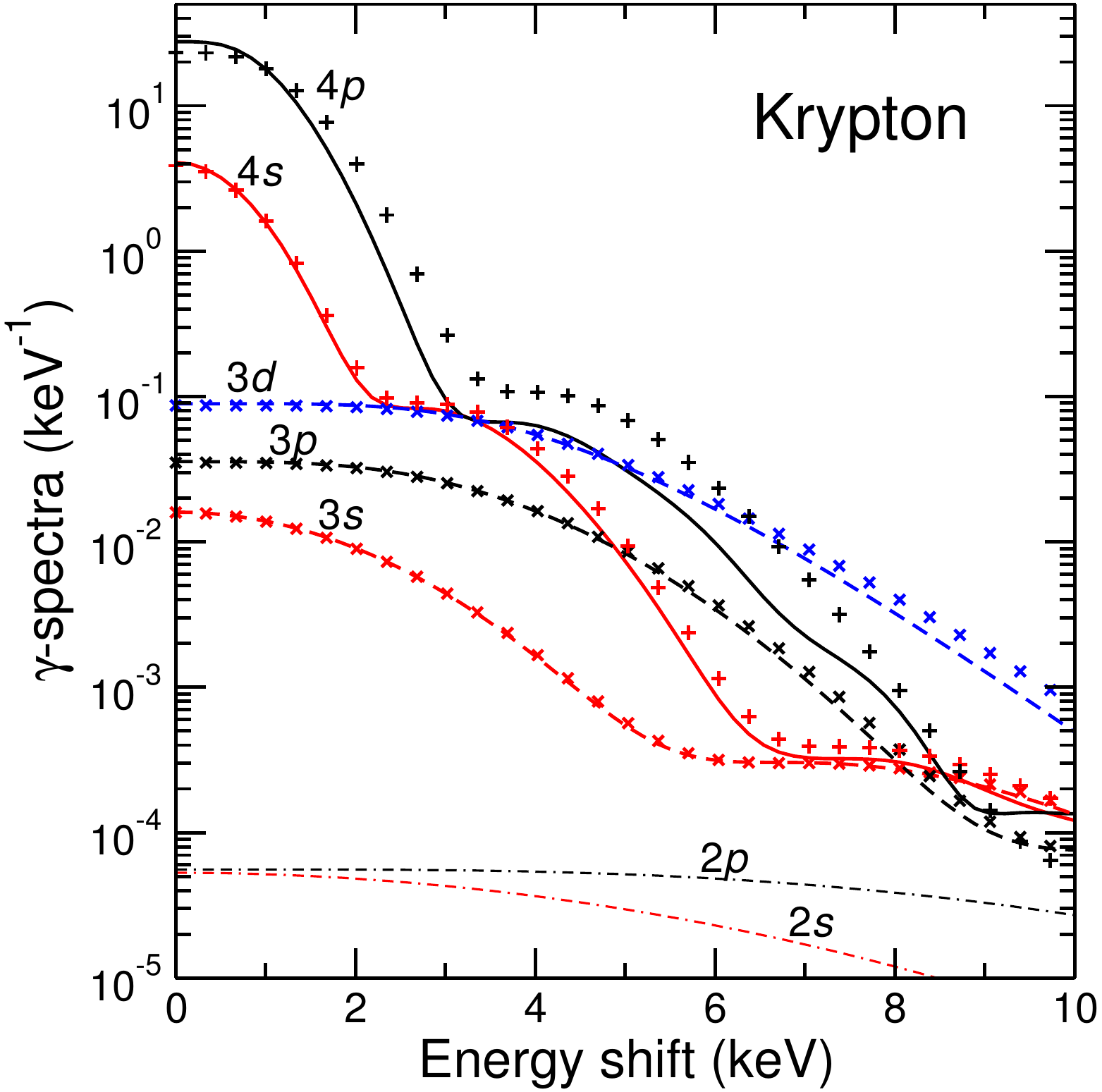}
\caption{$\gamma$-spectra for individual subshells of Krypton, calculated using the full many-body theory (dashed and solid lines for core and valence subshells respectively) and reconstructed from the IPA spectra (for Dyson positron) using the enhancement factor $\bar\gamma_{n}$ in accordance with \eqn{eqn:spectra_reconstructed} (crosses for core and plus symbols for valence).\label{fig:spectra_enhanced_kr}}
\end{figure}
For the subvalence (core) subshells the factors satisfy 1$<\bar\gamma_{\rm core}<$$\bar\gamma_{\rm valence}$, as one should expect. 
They range from 1.35 (Ar 2s) to 1.87 (Xe 4d). 
These values are appreciably different from unity, and highlight the necessary inclusion of correlations in the accurate calculation of the core, as well as the valence, $\gamma$-spectra.

Reconstruction of the true $\gamma$-spectra using these enhancement factors via \eqn{eqn:spectra_reconstructed} is demonstrated in \fig{fig:spectra_enhanced_kr} for the valence and subvalence subshells of Kr. 
Similar results are obtained for Ar and Xe \cite{DGG_thesis}. 
For the subvalence core (and the valence $4s$) subshells, the energy-independent enhancement factors are seen to successfully reconstruct the true spectra.
However, the reconstructed spectrum of the valence $4p$-subshell overestimates the full MBT result at large Doppler shifts and underestimates at small ones.
This is a result of the fact that both non-zero wave function density regions (`lobes') in the $4p$ radial wave function are scaled by the energy- and momentum- independent $\bar\gamma$ factor equivalently. 
Since the core enhancement is smaller than the valence, we expect the lobe at smaller radial distances, which contributes to the large energy (momentum) regions of the spectrum, to scale by less than the lobe at larger $r$, which contributes to the small energy regions of the spectrum. 

\begin{figure}[t!]
\includegraphics*[width=0.475\textwidth]{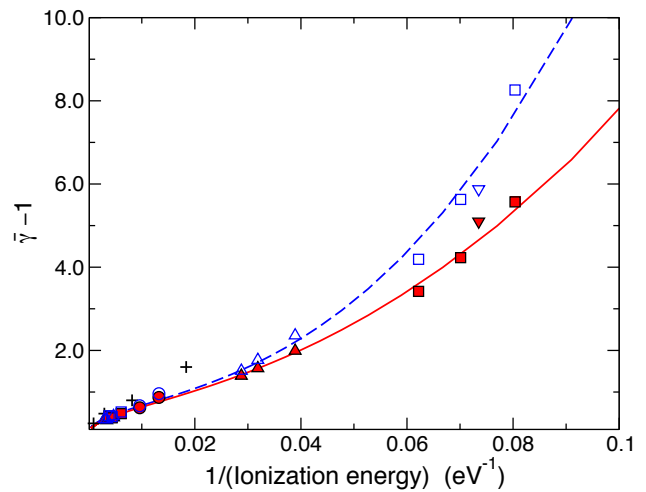}
\caption{Annihilation rate based enhancement factors $\bar{\gamma}_{n\varepsilon}$ for Ar (circles), Kr (squares) and Xe (diamonds), calculated using $s$-wave incident HF (solid symbols) and Dyson (unfilled symbols) positron.
Dashed line is the fit (\ref{eqn:enhancefit}) of $\gamma_{nl}$ for atoms obtained using the static HF positron wave function ($A = 42.0$ eV, $B = 24.9$ eV, $\beta = 2.54$), and the solid line is that for the Dyson positron wavefunction ($A = 35.7$ eV, $B = 22.7$ eV, $\beta$= 2.15).
\label{fig:enhancements}} 
\end{figure}
In spite of these complications for the outermost valence-shells, the simple enhancement factors work remarkably well for the core shells.
In \fig{fig:enhancements} we have plotted the enhancement factors of \tab{table:bargamma} against the inverse ionization energy of the subshell. 
The enhancement factor decreases in magnitude as the binding energy increases.

As we have seen, the vertex enhancement to the core spectra is dominated by the first-order correction; higher order corrections result in a minimal amount of further enhancement (see Fig.\,\ref{fig:spectra_npnm1p} and \tab{table:zeffs}).
One can estimate the strength of the short-range correlational enhancement from the first-order correction as the ratio $V_{\rm int}/\Delta E$, where $V_{\rm int}$ is the 
electron-positron Coulomb interaction potential, and $\Delta E$ is a measure of the 
response of the electrons to the perturbation (of the order of the respective electron's binding energy). 
For positron annihilation on H-like ions, this strength scales as $1/Z$, where $Z$ is the nuclear charge \cite{DGG_hlike}. 
For annihilation on the core electrons of many-electron atoms the strength can be thought to scale  analogously as $1/\sqrt{I_{E}}$, where $I_E$ is the ionization potential of the subshell.
One also arrives at a $1/\sqrt{I_E}$ dependence from simplifications to a calculation that explicitly includes two-particle Green's function of the electron-positron pair in the annihilation vertex \cite{DGG_thesis}.
In light of this, we have fitted the function
\begin{eqnarray}\label{eqn:enhancefit}
\bar\gamma_n-1\approx \sqrt\frac{A}{I_E}+ \left(\frac{B}{I_E}\right)^{\beta}
\end{eqnarray}
to the calculated enhancement factors of \tab{table:bargamma}, where $A$, $B$ and $\beta$ are constants. 
The first term describes the enhancement of the core electrons and is physically motivated as discussed above. 
The second term is phenomenological and is needed to describe the enhancement of the valence electrons, for which higher-order corrections are of as equal importance as the first-order one.
The solid line in \fig{fig:enhancements} is \eqn{eqn:enhancefit} fitted to the HF enhancement factors yielding the values $A=36$\,eV (1.32\au), $B=27.0$\,eV~(0.99\au), and $\beta=2.2$.
The dashed line is a fit of the function to the enhancement factors calculated with the Dyson incident positron wave function yielding the values $A=43.6$\,eV~(1.60\au), $B=20.5$\,eV~(0.75\au), and $\beta=2.6$. 
\begin{figure}[!t]
\includegraphics*[width=0.475\textwidth]{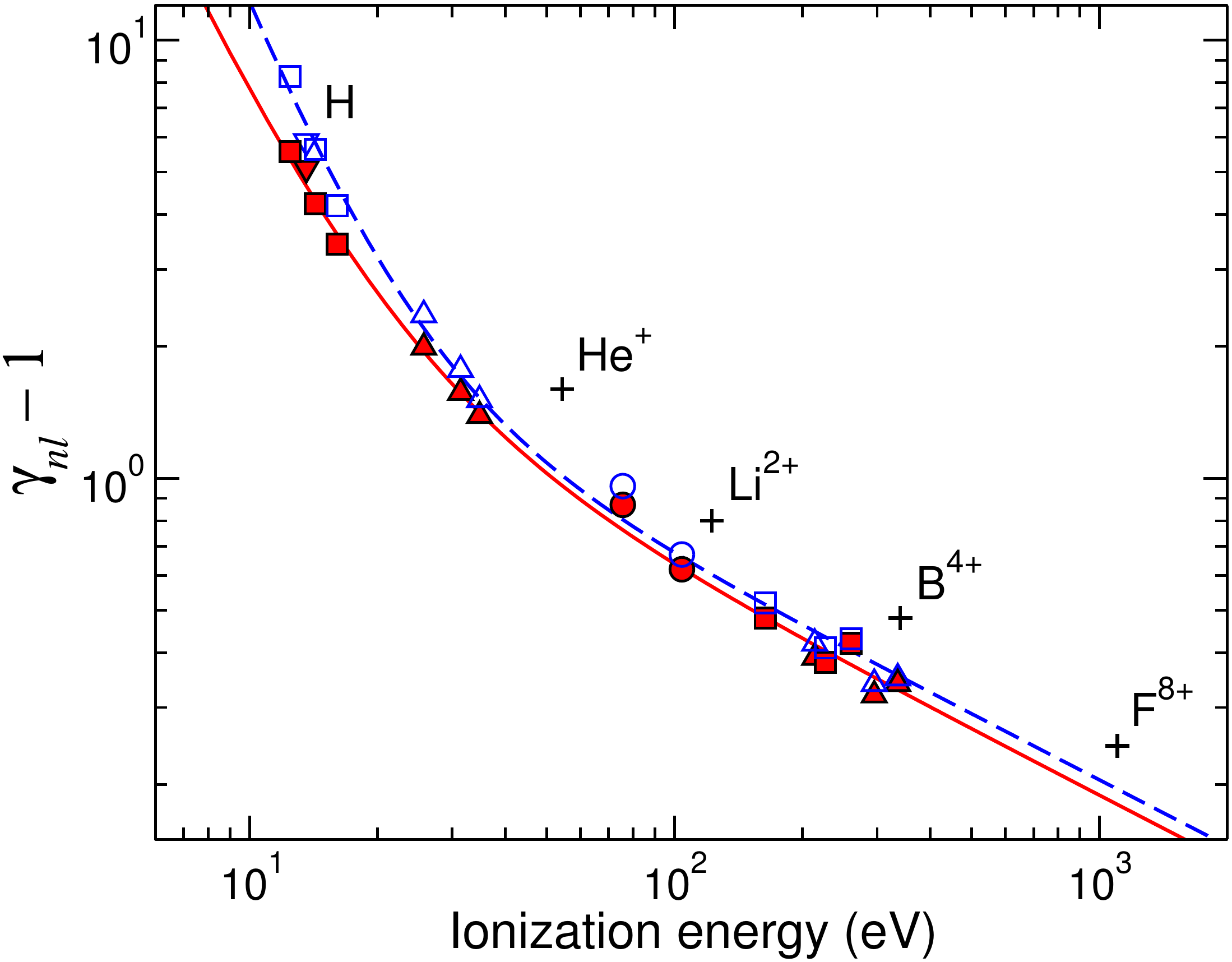}
\caption{
Enhancement factors (\ref{eqn:enhancefit}) calculated using static HF (open symbols) and Dyson (solid symbols) positron states, for the $s$ (triangles), $p$ (squares) and $d$ (circles) valence and core orbitals in Ar, Kr, and Xe; 1$s$ orbitals hydrogen (upside-down triangles) \cite{DGG_hlike}; and hydrogenlike ions (plus signs) \cite{DGG_hlike}. Dashed line is the fit (\ref{eqn:enhancefit}) of $\gamma_{nl}$ for atoms obtained using the static HF positron wave function ($A = 42.0$ eV, $B = 24.9$ eV, $\beta = 2.54$), and the solid line is that for the Dyson positron wavefunction ($A = 35.7$ eV, $B = 22.7$ eV, $\beta$= 2.15).
\label{fig:enhancementextrapolate}} 
\end{figure}

Finally, \fig{fig:enhancementextrapolate} shows the values of the enhancement factors plotted against the ionization energy of the subshell, as well as the functional form \eqn{eqn:enhancefit} extrapolated to large $I_E$. 
The $1/\sqrt{I_E}$ dependence for the core shells is evident. 
We also show the enhancements calculated for annihilation on Hydrogen and the H-like ions He$^+$, Li$^{2+}$, B$^{4+}$ and F$^{8+}$. 
Given the inherent difference between the positive ions and the neutral systems, the ions follow the trend surprisingly well.
For positrons annihilating in condensed matter systems, the longer-range positron-atom correlations included in the Dyson positron quasiparticle wave function average to zero. 
In this case the HF vertex enhancement factors, which neglect these longer-range type correlations in the positron wave function, would be expected to be more appropriate.

\section{Conclusion and outlook}\label{sec:conclusions}
$\gamma$-spectra for positron annihilation on the core and valence subshells of the noble gases Ar, Kr and Xe has been calculated using a many-body theory framework.
We have shown that the short-range electron-positron correlations, described through corrections to the zeroth-order (independent-particle-model) annihilation vertex, significantly enhance the $\gamma$-spectra magnitude and annihilation rate parameter $Z_{\rm eff}$. 
This enhancement is stronger for the valence shells, but it is still significant for the core shells (e.g., $\bar\gamma\simeq1.9$ for the $4d$ Xe subshell).
Despite the small probability of annihilation on the core subshells, we have shown that their contribution to the total annihilation $\gamma$-ray spectra is vital for accurate description of experiment across the full range of Doppler-shift energies, and particularly in the high-energy `wings'. 
We also use the atomic MBT to calculate \emph{true} vertex enhancement factors that take account of the non-local corrections to the annihilation vertex. We have shown that for a given subshell $n\ell$, these factors can be described by a simple and physically motivated function of the subshell ionization energy $I_{n\ell}$: $\bar{\gamma}_{n\ell}-1=\sqrt{A/I_{n\ell}}+(B/I_{n\ell})^{\beta}$, where $A$, $B$ and $\beta$ are positive constants. We have demonstrated the successful reconstruction of the core spectra using these enhancement factors to augment an independent-particle-model calculation. 
We suggest that this formula can be used to calculate the $\gamma$-ray spectra for positron annihilation on the core electrons of atoms across the periodic table, and for the atomic-like core electrons of condensed matter systems.
This work, taken together with Refs. \cite{DGG_posnobles} and \cite{DGG_coreprl}, gives a (near) complete description of the positron noble gas system.  

Looking to the future, there is still much work to be done to advance the theory and application of positron annihilation $\gamma$-spectra.
First, development and implementation of a fully relativistic many-body theory would shed light on the discrepancies between theory and experiment in the high-energy `wings' of the $\gamma$-spectra of Kr and Xe. 
Moreover, accurate predictions of positron binding and annihilation in complex molecules are required to develop new methods of vibrational spectroscopy \citep{RevModPhys.82.2557}, to help explain the origin of the strong $\gamma$-ray signal from the galactic centre \citep{RevModPhys.83.1001}, and are essential to realise \emph{spectroscopic} PET scanning \footnote{Existing PET scans use only the positional information, i.e., the origin of $\gamma$-ray production from a tracer that has reached a region of high-metabolic activity. In \emph{spectroscopic} PET scanning, the complementary detailed information in the energy spectra of the $\gamma$-rays would also be used, enabling non-invasive diagnostics.} and advance antimatter-matter chemistry.
Crucial in all cases is the ability of theory to accurately calculate the response of the atomic and molecular structure to the positron.  For molecules, (crude) preliminary calculations of $\gamma$-spectra for annihilation on small molecules, e.g., methane and its fluorosubstitutes, highlighted the strong effect of correlations and the possibility of using positron annihilation to probe electron momentum densities \citep{DGG_molgamma,DGG_molgammashort}. With some developments in the numerics, in particular an implementation of a Gaussian basis representation, accurate calculations for $\gamma$-spectra on molecules could be calculated with the existing theoretical framework.

\appendix
\section{Evaluation of $\gamma$-spectra from angular decomposition of the annihilation amplitude}\label{sec:appendix_angularreduction}
The diagrammatic expansion of the annihilation amplitude has been
given in \fig{fig:anndiags}. 
Its corresponding analytic form was given in \eqn{eqn:annamp_details}. It is a sum of three terms, 
each of which is in one-to-one correspondance with the diagrams in \fig{fig:anndiags}. 
Writing the amplitude in angular decomposed form (making use of graphical angular momenta techniques \cite{varshalovich}) and summing over the magnetic quantum numbers of the subshell $m_n$, one can evaluate the $\gamma$-spectrum as
\begin{eqnarray}\label{eqn:gammaspectra_reduced}
\bar{w}_{n\varepsilon}(\epsilon)=\frac{4}{ck}\int_{\frac{2|\epsilon|}{c}}^{\infty} \sum_{\lambda=|n-\varepsilon|}^{n+\epsilon}|A^{(\lambda)}_{n\varepsilon}(P)|^2PdP,
\end{eqnarray}
where
\begin{eqnarray}
A^{(\lambda)}_{n\varepsilon}(P)&\equiv \langle P||\hat{\bigotimes}^{(\lambda)}_{n\varepsilon}||n\varepsilon\rangle,
\end{eqnarray}
and the reduced vertex is given by
\begin{eqnarray}
\begin{split}
&\hat{\bigotimes}^{(\lambda)}_{n\varepsilon}
\equiv
\delta_{\lambda}-\sum_{\mu,\nu} \frac{\delta_{\lambda}||\mu\nu\rangle \langle \nu\mu ||V^{(\lambda)}}{E+\varepsilon_n-\varepsilon_{\mu}+\varepsilon_{\nu}}\\
&+\sum_{\mu_{i},\nu_{i}} \frac{\delta_{\lambda}||\mu_2\nu_2\rangle \langle \nu_2\mu_2 ||\Gamma^{(\lambda)}||\mu_1\nu_1 \rangle \langle \nu_1\mu_1 ||V^{(\lambda)}}{(E+\varepsilon_n-\varepsilon_{\mu_1}+\varepsilon_{\nu_1})(E+\varepsilon_n-\varepsilon_{\mu_2}+\varepsilon_{\nu_2})}
\end{split}
\end{eqnarray}
In this expression the reduced delta-function element (\emph{cf.} \eqn{eqn:defpmunu}) is defined by
\begin{eqnarray}
\begin{split}
{\langle P|| \delta_{\lambda} || 2, 1 \rangle}&\equiv 
\sqrt{[\ell_1][\lambda][\ell_2]}
\left(\begin{array}{ccc}
\ell_1 		&{\lambda}  		&\ell_2  \\
0 		& 0				& 0
\end{array}\right)\\
&\times\int j_{{\lambda}}(Pr)P_{\varepsilon_{1}\ell_1}(r)P_{\varepsilon_2 \ell_2}(r) dr.
\end{split}
\end{eqnarray}
where $j_{\lambda}$ is a spherical Bessel function, and we use the notation $[\ell]\equiv2\ell+1$.
The reduced Coulomb matrix element for a particle pair propagating with coupled angular momenta $\lambda$ is
\begin{eqnarray}
&&{\langle 3, 4 || V^{(\lambda)} || 2, 1 \rangle}
\equiv\sum_{l} (-1)^{{\lambda}+l} 
\left\{\begin{array}{ccc}{\lambda} & l_3 & l_4\\ l & l_2 & l_1 \end{array}\right\}
{\langle 3, 4|| V_{l} || 2, 1 \rangle},\nonumber \\
\end{eqnarray}
where the reduced Coulomb matrix
\begin{eqnarray}
\begin{split}
&{\langle 3, 4|| V_{l} || 2, 1 \rangle}\equiv \\
&\sqrt{\frac{[\ell_1][\ell_2][\ell_3][\ell_4]}{4\pi}}
\left(\begin{array}{ccc}
\ell_1 		&\ell  		&\ell_3  \\
0 		& 0		& 0
\end{array}\right)
\left(\begin{array}{ccc}
\ell_2 		&\ell 	&\ell_4  \\
0 		&0  	& 0 
\end{array}\right)\\
&\times\int P_{\varepsilon_{3}\ell_3}(r_1)P_{\varepsilon_{4}\ell_4}(r_2) \frac{r^\ell_<}{r^{\ell+1}_>}P_{\varepsilon_{2}\ell_2}(r_2)P_{\varepsilon_{1}\ell_1}(r_1) dr_1dr_2.
\end{split}
\end{eqnarray}

\section{Tabulated annihilation $\gamma$-spectra for the noble gases Ar, Kr and Xe}\label{sec:appendix_spectra}
\begin{table*}[htb!]
\caption{Annihilation $\gamma$-spectra $w(\varepsilon)$ for valence and subvalence
shells of the noble gases Ar, Kr and Xe. 
Results quoted are that of the full MBT calculation, i.e., fully-correlated Dyson incident positron and full annihilation vertex. 
Square brackets denote powers of ten. 
Spectra in different approximations to the incident positron wave function (HF or Dyson) and different orders of the annihilation vertex can be requested from the authors. \label{tab:spectra}}
\begin{ruledtabular}
\begin{tabular}{l@{\hspace*{-16pt}}c@{\hspace*{-16pt}}c@{\hspace*{-16pt}}c@{\hspace*{-16pt}}c@{\hspace*{-7pt}}c@{\hspace*{-16pt}}c@{\hspace*{-16pt}}c@{\hspace*{-16pt}}c@{\hspace*{-16pt}}c@{\hspace{-7pt}}c@{\hspace*{-16pt}}c@{\hspace*{-16pt}}c@{\hspace{-16pt}}c@{\hspace*{-16pt}}c@{\hspace*{-16pt}}c@{\hspace*{-16pt}}c@{\hspace*{-16pt}}c@{\hspace*{-16pt}}c@{\hspace*{-16pt}}cc@{\hspace*{-16pt}}}
		& \multicolumn{4}{c}{\hspace*{-24pt}Argon\hspace*{-24pt}	}	&  \multicolumn{5}{c}{\hspace*{-12pt}Krypton\hspace*{-24pt}}	 		& \multicolumn{5}{c}{\hspace*{-12pt}Xenon\hspace*{-26pt}}\\
$\epsilon$\,(keV)	& 2$s$ &2$p$ &$3s$ & $3p$	& 3$s$ & 3$p$ & $3d$ & $4s$ & $4p$	& 4$s$ & 4$p$ & $4d$ & $5s$ & $5p$\\ 
\hline
0.00	&	7.28[-3]	&	1.17	&	1.59	&	8.96	&	1.60[-2]	&	3.57[-2]	&	8.93[-2]	&	4.05	&	2.77[1]	&	9.15[-2]	&	2.44[-1]	&	9.38[-1]	&	2.14[1]	&	1.53[2]\\
0.25	&	7.24[-3]	&	1.17	&	1.51	&	8.92	&	1.59[-2]	&	3.57[-2]	&	8.93[-2]	&	3.79	&	2.74[1]	&	9.02[-2]	&	2.44[-1]	&	9.37[-1]	&	1.98[1]	&	1.51[2]\\
0.50	&	7.10[-3]	&	1.17	&	1.31	&	8.62	&	1.54[-2]	&	3.56[-2]	&	8.93[-2]	&	3.17	&	2.60[1]	&	8.63[-2]	&	2.44[-1]	&	9.37[-1]	&	1.56[1]	&	1.41[2]\\
0.75	&	6.88[-3]	&	1.17	&	1.03	&	7.83	&	1.48[-2]	&	3.56[-2]	&	8.92[-2]	&	2.38	&	2.27[1]	&	8.01[-2]	&	2.43[-1]	&	9.37[-1]	&	1.06[1]	&	1.17[2]\\
1.00	&	6.59[-3]	&	1.16	&	7.33[-1]	&	6.51	&	1.39[-2]	&	3.54[-2]	&	8.92[-2]	&	1.61	&	1.78[1]	&	7.22[-2]	&	2.40[-1]	&	9.35[-1]	&	6.11	&	8.29[1]\\
1.25	&	6.23[-3]	&	1.16	&	4.78[-1]	&	4.94	&	1.28[-2]	&	3.51[-2]	&	8.90[-2]	&	9.75[-1]	&	1.26[1]	&	6.32[-2]	&	2.35[-1]	&	9.30[-1]	&	3.08	&	5.02[1]\\
1.50	&	5.82[-3]	&	1.15	&	2.85[-1]	&	3.44	&	1.16[-2]	&	3.46[-2]	&	8.87[-2]	&	5.32[-1]	&	7.93	&	5.36[-2]	&	2.26[-1]	&	9.18[-1]	&	1.45	&	2.61[1]\\
1.75	&	5.36[-3]	&	1.14	&	1.57[-1]	&	2.21	&	1.04[-2]	&	3.38[-2]	&	8.80[-2]	&	2.70[-1]	&	4.49	&	4.41[-2]	&	2.14[-1]	&	8.97[-1]	&	7.96[-1]	&	1.18[1]\\
2.00	&	4.89[-3]	&	1.12	&	8.29[-2]	&	1.33	&	9.07[-3]	&	3.27[-2]	&	8.69[-2]	&	1.42[-1]	&	2.29	&	3.52[-2]	&	1.99[-1]	&	8.63[-1]	&	6.24[-1]	&	4.69\\
2.25	&	4.40[-3]	&	1.09	&	4.53[-2]	&	7.57[-1]	&	7.79[-3]	&	3.14[-2]	&	8.51[-2]	&	9.47[-2]	&	1.05	&	2.73[-2]	&	1.81[-1]	&	8.17[-1]	&	6.02[-1]	&	1.83\\
2.50	&	3.91[-3]	&	1.06	&	2.95[-2]	&	4.05[-1]	&	6.58[-3]	&	2.97[-2]	&	8.27[-2]	&	8.44[-2]	&	4.66[-1]	&	2.05[-2]	&	1.60[-1]	&	7.57[-1]	&	5.72[-1]	&	9.27[-1]\\
2.75	&	3.43[-3]	&	1.02	&	2.46[-2]	&	2.06[-1]	&	5.45[-3]	&	2.78[-2]	&	7.95[-2]	&	8.25[-2]	&	2.12[-1]	&	1.50[-2]	&	1.39[-1]	&	6.86[-1]	&	4.96[-1]	&	7.46[-1]\\
3.00	&	2.98[-3]	&	9.77	&	2.37[-2]	&	9.95[-2]	&	4.44[-3]	&	2.57[-2]	&	7.56[-2]	&	7.87[-2]	&	1.09[-1]	&	1.08[-2]	&	1.17[-1]	&	6.08[-1]	&	3.90[-1]	&	7.27[-1]\\
3.25	&	2.55[-3]	&	9.27	&	2.35[-2]	&	4.68[-2]	&	3.55[-3]	&	2.34[-2]	&	7.11[-2]	&	7.10[-2]	&	7.52[-2]	&	7.70[-3]	&	9.64[-2]	&	5.26[-1]	&	2.80[-1]	&	6.84[-1]\\
3.50	&	2.16[-3]	&	8.72	&	2.24[-2]	&	2.30[-2]	&	2.79[-3]	&	2.11[-2]	&	6.60[-2]	&	6.02[-2]	&	6.75[-2]	&	5.58[-3]	&	7.72[-2]	&	4.44[-1]	&	1.85[-1]	&	5.91[-1]\\
3.75	&	1.81[-3]	&	8.15	&	2.04[-2]	&	1.34[-2]	&	2.15[-3]	&	1.87[-2]	&	6.05[-2]	&	4.81[-2]	&	6.57[-2]	&	4.24[-3]	&	6.02[-2]	&	3.66[-1]	&	1.13[-1]	&	4.72[-1]\\
4.00	&	1.49[-3]	&	7.55	&	1.77[-2]	&	1.02[-2]	&	1.64[-3]	&	1.64[-2]	&	5.47[-2]	&	3.64[-2]	&	6.26[-2]	&	3.46[-3]	&	4.57[-2]	&	2.95[-1]	&	6.44[-2]	&	3.53[-1]\\
4.25	&	1.22[-3]	&	6.93	&	1.47[-2]	&	9.55[-3]	&	1.24[-3]	&	1.41[-2]	&	4.90[-2]	&	2.62[-2]	&	5.67[-2]	&	3.08[-3]	&	3.38[-2]	&	2.33[-1]	&	3.46[-2]	&	2.49[-1]\\
4.50	&	9.83[-4]	&	6.32	&	1.18[-2]	&	9.48[-3]	&	9.36[-4]	&	1.20[-2]	&	4.33[-2]	&	1.79[-2]	&	4.85[-2]	&	2.92[-3]	&	2.43[-2]	&	1.80[-1]	&	1.81[-2]	&	1.66[-1]\\
4.75	&	7.83[-4]	&	5.72	&	9.11[-3]	&	9.30[-3]	&	7.11[-4]	&	1.01[-2]	&	3.79[-2]	&	1.18[-2]	&	3.96[-2]	&	2.88[-3]	&	1.70[-2]	&	1.35[-1]	&	9.98[-3]	&	1.04[-1]\\
5.00	&	6.17[-4]	&	5.13	&	6.86[-3]	&	8.85[-3]	&	5.52[-4]	&	8.39[-3]	&	3.28[-2]	&	7.57[-3]	&	3.16[-2]	&	2.87[-3]	&	1.17[-2]	&	1.00[-1]	&	6.55[-3]	&	6.31[-2]\\
5.25	&	4.81[-4]	&	4.58	&	5.03[-3]	&	8.17[-3]	&	4.45[-4]	&	6.86[-3]	&	2.81[-2]	&	4.69[-3]	&	2.48[-2]	&	2.85[-3]	&	7.94[-3]	&	7.23[-2]	&	5.39[-3]	&	3.91[-2]\\
5.50	&	3.72[-4]	&	4.05	&	3.59[-3]	&	7.31[-3]	&	3.77[-4]	&	5.54[-3]	&	2.39[-2]	&	2.80[-3]	&	1.90[-2]	&	2.77[-3]	&	5.42[-3]	&	5.12[-2]	&	5.14[-3]	&	2.50[-2]\\
5.75	&	2.85[-4]	&	3.56	&	2.51[-3]	&	6.32[-3]	&	3.37[-4]	&	4.41[-3]	&	2.01[-2]	&	1.64[-3]	&	1.42[-2]	&	2.64[-3]	&	3.81[-3]	&	3.54[-2]	&	5.09[-3]	&	1.58[-2]\\
6.00	&	2.18[-4]	&	3.11	&	1.72[-3]	&	5.29[-3]	&	3.16[-4]	&	3.47[-3]	&	1.68[-2]	&	9.64[-4]	&	1.01[-2]	&	2.46[-3]	&	2.87[-3]	&	2.40[-2]	&	4.94[-3]	&	9.41[-3]\\
6.25	&	1.68[-4]	&	2.70	&	1.16[-3]	&	4.32[-3]	&	3.07[-4]	&	2.69[-3]	&	1.40[-2]	&	6.05[-4]	&	6.92[-3]	&	2.24[-3]	&	2.36[-3]	&	1.59[-2]	&	4.57[-3]	&	5.35[-3]\\
6.50	&	1.30[-4]	&	2.33	&	7.67[-4]	&	3.51[-3]	&	3.05[-4]	&	2.06[-3]	&	1.15[-2]	&	4.33[-4]	&	4.63[-3]	&	2.00[-3]	&	2.12[-3]	&	1.03[-2]	&	4.02[-3]	&	3.39[-3]\\
6.75	&	1.03[-4]	&	2.00	&	4.99[-4]	&	2.88[-3]	&	3.04[-4]	&	1.56[-3]	&	9.44[-3]	&	3.59[-4]	&	3.18[-3]	&	1.74[-3]	&	2.03[-3]	&	6.54[-3]	&	3.42[-3]	&	2.84[-3]\\
7.00	&	8.34[-5]	&	1.71	&	3.18[-4]	&	2.39[-3]	&	3.03[-4]	&	1.16[-3]	&	7.69[-3]	&	3.33[-4]	&	2.33[-3]	&	1.49[-3]	&	2.01[-3]	&	4.10[-3]	&	2.85[-3]	&	2.80[-3]\\
7.25	&	7.05[-5]	&	1.46	&	1.98[-4]	&	2.00[-3]	&	3.01[-4]	&	8.60[-4]	&	6.24[-3]	&	3.27[-4]	&	1.83[-3]	&	1.24[-3]	&	2.01[-3]	&	2.56[-3]	&	2.37[-3]	&	2.75[-3]\\
7.50	&	6.22[-5]	&	1.23	&	1.22[-4]	&	1.65[-3]	&	2.95[-4]	&	6.30[-4]	&	5.03[-3]	&	3.25[-4]	&	1.45[-3]	&	1.02[-3]	&	1.99[-3]	&	1.63[-3]	&	1.98[-3]	&	2.66[-3]\\
7.75	&	5.71[-5]	&	1.04	&	7.69[-5]	&	1.32[-3]	&	2.86[-4]	&	4.60[-4]	&	4.04[-3]	&	3.21[-4]	&	1.11[-3]	&	8.24[-4]	&	1.94[-3]	&	1.11[-3]	&	1.64[-3]	&	2.61[-3]\\
8.00	&	5.44[-5]	&	8.71	&	5.25[-5]	&	1.02[-3]	&	2.74[-4]	&	3.36[-4]	&	3.24[-3]	&	3.10[-4]	&	7.94[-4]	&	6.54[-4]	&	1.86[-3]	&	8.28[-4]	&	1.33[-3]	&	2.61[-3]\\
8.25	&	5.30[-5]	&	7.28	&	4.08[-5]	&	7.54[-4]	&	2.60[-4]	&	2.49[-4]	&	2.58[-3]	&	2.90[-4]	&	5.29[-4]	&	5.09[-4]	&	1.75[-3]	&	6.94[-4]	&	1.04[-3]	&	2.62[-3]\\
8.50	&	5.25[-5]	&	6.06	&	3.60[-5]	&	5.46[-4]	&	2.43[-4]	&	1.88[-4]	&	2.06[-3]	&	2.63[-4]	&	3.38[-4]	&	3.90[-4]	&	1.62[-3]	&	6.40[-4]	&	7.60[-4]	&	2.63[-3]\\
8.75	&	5.24[-5]	&	5.03	&	3.44[-5]	&	4.02[-4]	&	2.26[-4]	&	1.48[-4]	&	1.63[-3]	&	2.32[-4]	&	2.29[-4]	&	2.94[-4]	&	1.47[-3]	&	6.24[-4]	&	5.21[-4]	&	2.64[-3]\\
9.00	&	5.23[-5]	&	4.16	&	3.40[-5]	&	3.14[-4]	&	2.07[-4]	&	1.23[-4]	&	1.29[-3]	&	2.02[-4]	&	1.87[-4]	&	2.19[-4]	&	1.31[-3]	&	6.21[-4]	&	3.36[-4]	&	2.55[-3]\\ 
\end{tabular}
\end{ruledtabular}
\end{table*}

\begin{acknowledgments}
We thank Prof.~C.~M.~Surko for valuable discussions.
\end{acknowledgments}

%

\end{document}